\documentclass[]{emulateapj}
\usepackage{amsmath}
\usepackage{natbib}
\slugcomment{Draft version}

\shorttitle{A simple model linking galaxy and dark matter evolution}
\shortauthors{Birrer et al.}

\begin{document}

\title{A simple model linking galaxy and dark matter evolution}

\author{Simon Birrer}
 \email{simon.birrer@phys.ethz.ch}
\author{Simon Lilly}
 \email{simon.lilly@phys.ethz.ch}
\author{Adam Amara}
\author{Aseem Paranjape}
\author{Alexandre Refregier}
\affiliation{%
Institute for Astronomy, Department of Physics, ETH Zurich, Wolfgang-Pauli-Strasse 27, 8093, Zurich, Switzerland
}%

\date{\today}

\begin{abstract}
We construct a simple phenomenological model for the evolving galaxy population by incorporating pre-defined baryonic prescriptions into a dark matter hierarchical merger tree. Specifically the model is based on the simple gas-regulator model introduced by \cite{Lilly:2013p7455} coupled with the empirical quenching rules of \cite{Peng:2010p132, Peng:2012p1015}.
The simplest model already does quite well in reproducing, without re-adjusting the input parameters, many observables including the Main Sequence sSFR-mass relation, the faint end slope of the galaxy mass function and the shape of the star-forming and passive mass functions. Compared with observations and/or the recent phenomenological model of \cite{Behroozi:2013p7464} based on epoch-dependent abundance-matching, our model also qualitatively reproduces the evolution of the Main Sequence sSFR(z) and SFRD(z) star formation rate density relations, the $M_s - M_h$ stellar-to-halo mass relation and also the $SFR - M_h$ relation. Quantitatively the evolution of sSFR(z) and SFRD(z) is not steep enough, the $M_s - M_h$ relation is not quite peaked enough and, surprisingly, the ratio of quenched to star-forming galaxies around M* is not quite high enough. We show that these deficiencies can simultaneously be solved by ad hoc allowing galaxies to re-ingest some of the gas previously expelled in winds, provided that this is done in a mass-dependent and epoch-dependent way. These allow the model galaxies to reduce an inherent tendency to saturate their star-formation efficiency. This emphasizes how efficient galaxies around M* are in converting baryons into stars and highlights the fact that quenching occurs just at the point when galaxies are rapidly approaching the maximum possible efficiency of converting baryons into stars.

\end{abstract}


\maketitle

\section{Introduction}
Galaxy evolution is a field where cosmological structure formation needs to be enriched with astrophysical processes, i.e. astrophysics has to be embedded into a cosmological model. It is the largest scale where astrophysical models have to succeed and the smallest scales where the cosmological structure formation model has to prove its validity. Galaxies, and the galaxy population, therefore offer tests for both astrophysics and cosmology. 
 
Several approaches have been taken to understand the link between galaxies and dark matter haloes. Usually, the dark matter component is assumed to be well understood on the basis of both analytic and numerical models that are based on input parameters derived from cosmological observations, e.g. the cosmic microwave background. Small collapsed objects, i.e. "haloes", form earlier and subsequently merge together to form more massive objects. Numerical N-body simulations provide an accurate description of the evolution of the population of dark matter haloes in the cosmological context \citep[e.g,][]{Springel:2003p1375, Klypin:2011p3233}. Much of the difficulty in galaxy formation and evolution arises then in understanding the actions of baryonic physics within these haloes.

A major theoretical effort has been made using so-called 'semi-analytic' techniques to follow the evolution of baryons in the haloes. In semi-analytic models (or SAMs) simple parametric descriptions of the most important baryonic physics are combined with a dark matter merger tree that is usually obtained from a large volume N-body simulation. The treatment of the relevant baryonic processes is necessarily simplified \citep[e.g,][]{Lacey:1991p1385,White:1991p1386,Kauffmann:1993p5842,Somerville:1999p5853,Kauffmann:1999p5861,Springel:2001p5885,Helly:2003p5900,Hatton:2003p5907,Springel:2005p331}. Some or all of the parameters describing these processes can be adjusted to match particular observational properties of galaxies or of the galaxy population, either at a single epoch or at many. Although much progress has been made and the range of output quantities can be large, the total number of parameters in such models is often quite large and as a result, the uniqueness and predictive power of SAMs is limited. In addition, the apparent complexity of the SAMs can often hide underlying links between different aspects of galaxy evolution.

Much progress has also been made using the alternative approach of ab initio simulations in which the baryonic physics is directly incorporated into hydrodynamic codes. However, due to the very large dynamical range that must be covered, such simulations are currently not able to resolve star formation and associated feedback processes and so cannot describe these processes from first principles. Simulation codes therefore include these as ''sub-grid" physics, which leads to the emergence of a number of alternative approaches \citep[e.g,][]{Springel:2005p331,Croton:2006p3228}.

Partly in response to these difficulties, other, more phenomenological, approaches have been developed.
One has been to study the statistical connection between galaxies and dark matter haloes in terms of the conditional luminosity function \citep[CLF;][]{Yang:2003p276} or the halo occupation distribution \citep[HOD; e.g,][]{Peacock:2000p239, Seljak:2000p3239}. These methods are anchored on our good understanding of the statistical properties of dark matter haloes in the current $\Lambda$CDM model plus the hypothesis that galaxy properties should be closely linked to the properties (and especially the masses) of dark matter haloes. A variety of statistical tools can then be used to constrain the galaxy-dark matter connection: galaxy clustering \citep[e.g,][]{Zehavi:2011p3244}, galaxy-galaxy lensing \citep[e.g,][]{Brainerd:1996p3343, Sheldon:2004p3421, Leauthaud:2010p3441}, galaxy group catalogs \citep[e.g,][]{Berlind:2006p3981, Yang:2007p3863}, abundance matching \citep[recently e.g,][]{Leauthaud:2012p7229, Hearin:2013p3511, Reddick:2013p3450, Tinker:2013p7227}, and satellite kinematics \citep[e.g,][]{More:2009p3924}. 

In recent years, large scale surveys of the distant Universe have yielded sufficient data to apply similar approaches at significant look-back times. The differential effects with redshift then allow a phenomenological description of the evolving galaxy population using simple parametric descriptions. The parameters of these are matched to the evolving statistical descriptions of the stellar to halo mass relation \citep[e.g,][]{Firmani:2010p7360,Yang:2012p243,Behroozi:2013p7464,Lu:2014p7573}. Such models can provide consistency checks within several data sets and observables. As an example, when compiling different data sets, \cite{Behroozi:2013p7464} finds a disagreement between galaxy abundances for high redshift surveys and high systematic errors in the stellar mass and star formation rate estimates.

The increasingly good observational data on the evolving galaxy population has also opened up other phenomenological approaches which instead focus on the baryonic processes. A successful approach has been to broadly classify galaxies as either actively forming stars or quiescent. Most star-forming galaxies exhibit a rather tight relation between their star formation rates (SFR) and stellar masses producing the so called Main Sequence \citep[]{Brinchmann:2004p5336, Noeske:2007p706, Daddi:2007p837, Peng:2010p132, Rodighiero:2011p1606}. The quiescent galaxies have sSFR that are 1-2 orders of magnitude lower, and these galaxies are not forming stars at a cosmologicaly significant rate. We will henceforth refer to these passive galaxies as ``quenched''. A few underlying simplicities in the galaxy population can then be identified (such as the observed constancy of the Schechter $M^*$ of star-forming galaxies or the separability of the fraction of galaxies that are quenched - colloquially the ``red fraction''). The analytic consequences of these can then be explored using the most basic continuity equations \citep[][hereafter P10 and P12]{Peng:2010p132,Peng:2012p1015}. This has proved very successful in describing the evolution of the galaxy population and, in particular, in deriving the simple empirical "laws" for the quenching of star-formation in galaxies as a function of stellar mass (even if other parameters are involved or are even the main causal drivers). This approach has also yielded new insights into the relationships between the mass functions of active and passive galaxies, and the relative importance of mass and environment in the quenching of star-formation in galaxies. 

There have also been several papers developing simple toy analytic models for the star-formation rate in galaxies \citep[e.g,][L13 from here on]{Bouche:2010p311,Dave:2011p81,Krumholz:2012p5540,Dekel:2014p5804,Dayal:2013p5951,Lilly:2013p7455}. These have been motivated by the small dispersion in the specific star-formation rate (sSFR = star formation rate/stellar mass) of actively star-forming galaxies, and by the strong evolution of this characteristic sSFR with time. In terms of the CLF a phenomenological approach has been choosen by \cite{Tacchella:2013p416}. \cite{Dekel:2013p5802} developed a toy analytic model when comparing to hydrodynamical simulations. All these models have tried to boil down the complexity arising in numerical simulations and detailed semi-analytic models into simple analytic models that are motivated by either simulation results or observational constraints. The aim has been to provide a simple picture of how galaxies evolve in the cosmological context and to highlight connections between different aspects of galaxy evolution. In particular, L13 developed a toy analytic model in which the star-formation rate is regulated via the variable mass of gas in the gas reservoir feeding the star-formation. Such a model links the specific star-formation rate (sSFR) to the specific accretion rate onto the regulator system. The self-regulation by the gas reservoir naturally introduces the SFR as a second parameter in the mass-metallicity relation $Z(m,\text{SFR})$ and also naturally explains why the $Z(m,\text{SFR})$ relation should be more or less independent of time. This model also links in a straightforward way the different slopes of the mass functions of galaxies and haloes.

By construction, the phenomenological analytic models in P10, P12 and L13 have only been tangentially linked to the dark matter haloes and not at all to the overall evolving {\it population} of haloes that is produced by hierarchical assembly in the cosmological context. The whole approach, and in particular the derivation of the numerical values of the few parameters in the models, was based on comparison with baryonic systems. This has been both a strength and a weakness of these analyses.

The aim of this paper is therefore to explore how far we can get by taking these simple baryonic prescriptions and combine them with a dark matter structure formation formalism. Specifically, we will take the self-regulation model of L13 plus the quenching "laws" of P10 and P12. We couple them with a Monte Carlo realisation of dark matter halo merger trees. Our goal is to present a phenomenological model whose few parameters are taken from the earlier papers, and are not adjusted in the combined model. The parameters are well constrained and therefore considered as non-adjustable in this paper. We can then explore how well these predictions match the observed Universe, and identify where and how it needs further improvement. In a second step, we propose two changes in the model and show their impact on the predictions.

Our approach is thus rather different to the one in \cite{Lu:2014p7573} or \cite{Behroozi:2013p7464} as we do not explore a parameter space but rather develop a physical picture without further tuning within the combined model. We stress that the current model is not intended to replace more complex SAMs whose greater sophistication will no doubt be required to account for a more multi-dimensional view of galaxies.

The current paper is structured as follow: In Section \ref{sec:concepts} we review the key concepts that were introduced in the earlier papers P10, P12 and L13 which we use to establish the characteristics of the baryonic processes. We define our notation and parameterization of these independent models and describe the dark matter structure formation formalism we apply. In Section \ref{sec:combined} we describe how these are then combined into the dark matter merger tree, and what further assumptions have to be added, and how the model is then run. In Section \ref{sec:results}, we present our results in terms of the most basic observables of the galaxy population such as the overall star-formation rate density (SFRD), the sSFR-mass relation of star-forming galaxies, the mass function of active and passive galaxies, and the form of the stellar mass vs. halo mass relation for star-forming and passive galaxies, and compare them with other work. In Section \ref{sec:discussion} we discuss the implications of the model and explore how one could modify it and where we are more restricted by the linkages between different parts of the model. Finally, in Section \ref{sec:conclusion} we summarize our conclusions.

Throughout this paper, we assume a flat cosmology with $h=0.7$ (i.e. $H_0 = 70$kms$^{-1}$Mpc$^{-1}$), $\Omega_b=0.045$, $\Omega_m=0.3$, $\Omega_{\lambda}=0.7$ , $\sigma_8=0.8$ and $n_s=1.0$ consistent with \cite{Komatsu:2011p966} WMAP7 results. We use the BBKS \citep[][]{Bardeen:1986p980} transfer function to calculate the matter power spectrum. We define a halo as having a mean over-density $\Delta \equiv 3M_h/4\pi \Omega_m \rho_{\text{crit}}R_h^3=170$ to be consistent with the merger tree we use in this paper.
We use "dex" to refer to the anti-logarithm, so that 0.3 dex represents a factor of 2.

\section{Model ingredients}

\label{sec:concepts}

In this section we review the concepts and descriptions used in our model. We start with the differential equations that control the regulator system from L13 (Section \ref{sec:regulator}). We then quote the mass- and satellite- quenching expressions from P10 and P12 (Section \ref{sec:quenching}). In Section \ref{sec:dm-process} we describe the dark matter structure formation formalism we apply to our model. These ingredients are completely independent of each other and do not rely on mechanisms described in other subsections.

\subsection{Galaxies as gas-regulated systems}\label{sec:regulator}

We adopt the model proposed in L13. Several similar models have been proposed in the literature \citep[e.g,][]{Bouche:2010p311,Dave:2011p81,Krumholz:2012p5540,Dekel:2014p5804,Dayal:2013p5951} althought there are significal differences in both concept and detail. We identify a galaxy as a gas-regulated system sitting in a dark matter halo. The SFR in the galaxy is set simply by the gas mass $M_{gas}$ within a reservoir in the galaxy via a star-formation efficiency, $\epsilon$. There is also mass-loss from the reservoir in the form of a wind that is parameterized by a mass-loading factor, $\lambda$, such that the outflow is $\lambda \cdot$SFR. Both of the $\epsilon$ and $\lambda$ parameters are allowed to vary with the stellar mass $M_s$ of the galaxy (and possibly the epoch, or redshift). In L13, the baryonic infall rate into the regulator $\Phi_b$, which replenishes the reservoir, was assumed to be some fixed fraction $f_{\text{gal}}$ of the baryonic infall onto the surrounding halo. Two obvious simplifications of the L13 model were that gas expelled from the galaxy in the wind was assumed to be lost forever, i.e. it does not mix with any surrounding gas in the halo, and that substructure within a halo was neglected, i.e. there was only one regulator in each halo. These issues will be discussed later in this paper.

As in L13, the stellar mass is defined as the long lived stellar population assuming that a fraction $R$ of newly formed stellar mass is promptly returned to the gas reservoir. The remaining stars will have a lifetime that is longer than the Universe. As in L13, we will set the mass-return factor $R=0.4$, motivated by stellar population models \citep[e.g,][]{Bruzual:2003p700}. The "stellar masses" used throughout this paper will be these "long-lived" stellar masses. These are of order 0.2 dex smaller than the stellar masses that are obtained by integrating the SFR, which are sometimes quoted in the literature.

The build up in stellar mass $\dot{M}_s$ is then given by 
\begin{equation}
\label{eqn:reduced}
	\dot{M}_s=SFR\cdot (1-R).
\end{equation}
Following L13, the differential equations of the regulator in differential form can then be written as:
\begin{equation}
\label{eqn:m_s_dot}
	SFR=\epsilon \cdot M_{\text{gas}}
\end{equation}
\begin{equation}
	\dot{M}_{\text{gas,outflow}}=\lambda \cdot SFR
\end{equation}
\begin{equation}
	\dot{M}_{\text{gas}}=\Phi_b - \dot{M}_s - \dot{M}_{\text{gas,outflow}}=\Phi_b -\epsilon \left(1-R+\lambda \right)M_{\text{gas}}
\end{equation}
We will not go in detail into the analytic solution of these differential equations as L13 explored these in some detail.

The efficiency $\epsilon$ and the outflow load $\lambda$ are intended to cover, albeit simplistically, all the baryonic processes within the galaxy system. L13 considered a power law parametrization for both these quantities as a function of the stellar mass $M_s$  in order to match the observed Z($M_s$,SFR) relation in \cite{Mannucci:2010p318}. The parameterization as a function of stellar mass is a convenience and is still valid even if other quantities (e.g. halo mass) are responible for the physical effect. The parameterization is: 
\begin{equation}
\label{eqn:epsilon}
	\epsilon(M_s,z)=\epsilon_{10}\cdot \left(\frac{M_s}{10^{10}M_{\odot}}\right)^b\cdot \left( \frac{H(z)}{H_0}\right)
\end{equation}
\begin{equation}
\label{eqn:lambda}
	\lambda(M_s)=\lambda_{10}\cdot \left(\frac{M_s}{10^{10}M_{\odot}}\right)^a,
\end{equation}

$H(z)$ is the Hubble rate at redshift $z$ and $H_0$ the present-day Hubble constant. L13 assumed, following \cite{Mo:1998p1153}, that the star-formation efficiency would scale as the inverse dynamical time of the galaxies and haloes, which should scale as the Hubble rate, and we will do the same until revisiting this issue towards the end of the paper\footnote[1]{The actual redshift scaling of the efficiency in L13 is $\epsilon \propto (1+z)$ which is a good approximation for the scaling of the Hubble rate at low redshifts.}. For example \cite{Feldmann:2013p5526} looked at the the role of the normalization and slope of the Kennicutt-Schmidt relation (our Equation \ref{eqn:m_s_dot}) by varying this parameter and keeping all other parameters fixed. They find that a linear Kennicutt-Schmidt relation is a much better fit to observations than a strongly super-linear relation, in agreement with L13.

The gas infall rate $\Phi_b$ is assumed to be closely related to the dark matter halo growth rate. We will describe this term in greater detail when discussing our model in Section \ref{sec:combined} but in essence we set the $f_{\text{gal}}$ parameter of L13 to unity, i.e. all of the gas flowing in a halo will be assigned (at least temporary) with a regulator system.

One of the most interesting features of this very simple regulator system is that the resulting sSFR is closely linked to the specific infall rate of the baryons, which L13 termed the $SMIR_B$. 

Indeed, the model is motivated by the overall similarities between the observed sSFR$(z)$ of the population of star-forming galaxies and the specific growth rate of dark matter haloes \citep[e.g,][or L13]{Schaye:2010p5795}. The sSFR will be exactly the specific baryonic infall rate if a constant fraction $f_{\text{star}}$ of the incoming gas is converted into stars. If, however, this fraction increases as a given regulator evolves, e.g. if star-formation becomes more efficient as the stellar mass of the regulator increases, then the sSFR will be boosted relative to the specific baryon infall rate, as in Equation 36 of L13. Because this boosting of the sSFR is likely to be larger at low masses, this also has the effect of reversing the weak dependence of the sSFR on stellar mass relative to the dependence of the dark matter specific accretion rate on halo mass (see L13).

Another attractive feature of this regulator system is that it introduces the SFR as a second parameter in the mass-metallicity relation, producing a $Z(M_{s},\text{SFR})$ relation that will only change with epoch to the extent that the $\epsilon$ and $\lambda$ parameters (at fixed $M_{s}$) evolve. In other words a so-called "fundamental metallicity relation" is a more-or-less natural outcome of the regulator. By comparing the expected $Z(M_{s},\text{SFR})$ with data from SDSS given by \cite{Mannucci:2010p318}, L13 derived nominal values for the parameters $\epsilon_{10}$, $b$, $\lambda_{10}$ and $a$ in Equation (\ref{eqn:epsilon}) and (\ref{eqn:lambda}) above. Given the extreme simplicity of the model, the resulting values for $\epsilon(M_{s})$ and $\lambda(M_{s})$, which are quoted in Table 1 in L13 and included in Table \ref{tab:parameters} of this paper, are surprisingly reasonable, giving gas depletion timescales ($\epsilon^{-1}$) at $M_{s} \sim 10^{10}$M$_{\odot}$ of about 2 Gyr and mass-loading factors of order unity. The gas depletion timescale and the outflow mass loading both decrease with increasing stellar mass resulting in more and more efficient conversion of inflowing baryons into stars as the stellar mass of the system increases. The fraction of incoming baryons that are converted to stars is denoted as $f_{\text{star}}$ in L13. In the context of the simple analysis of L13, this ''saturation'' of $f_{\text{star}}$ can be traced to the pronounced flattening of the $Z(M_{s})$ relation at high masses. We will return to this later in the paper.

The processes associated with star-formation in galaxies are thus represented in our model by the four parameters (Equation \ref{eqn:epsilon} and \ref{eqn:lambda}) describing $\epsilon(M_{s})$ and $\lambda(M_{s})$, and taken straight from L13. As noted above, we will initially assume $\epsilon$ increases as $H(z)/H_0$, although we will revisit this assumption later.

Work by \citep[e.g,][]{Springel:2003p5955,Benson:2003p5972,DeLucia:2004p6956,Governato:2007p6057,Oppenheimer:2008p6820,Scannapieco:2008p6281,Bower:2012p6284} have emphasized the importance of supernova feedback. In L13, outflows of material represent an "inefficnecy" in the production of stars, but do not "regulate" the level of star-formation, which is instead defined by the gas mass.

\subsection{Quenching of star-formation in galaxies}

\label{sec:quenching}

In this paper, we apply the phenomenological quenching prescriptions derived by P10 and P12. This is distinct from introducing a turnover in the efficiency parameter as done by \cite{Behroozi:2013p7464} and \cite{Lu:2014p7573} or cutting off the supply of gas, as done by e.g. \cite{Bouche:2010p311}, although the outcomes may be similar. There are many physical mechanisms that have been proposed for quenching. One popular approach is AGN feedback \citep[see e.g,][]{Governato:2004p6576,Croton:2006p3228,Bower:2006p7156,Booth:2009p6622}. The AGN feedback also presents a viable solution to the cooling flow problem \citep[see e.g,][]{Fabian:1994p6645,Bohringer:2002p6713,Ishibashi:2012p6663}, hence its popularity. The P10 approach comes from the continuity in the two distinct galaxy populations and is not based on a particular physical mechanism but rather seeks to define the characteristics that any viable mechanism must satisfy.

We will assume that star formation within a galaxy stops instantaneously when it is quenched and that no significant star formation occurs afterwards. As a shorthand (and on plots) we will denote the actively star-forming galaxies as blue and those that are quenched as red although we will not consider the colors of galaxies per se. The ``red fraction'' will then be the fraction of galaxies of a given mass etc. that have been quenched.

P10 showed that the red fraction of galaxies as a function of mass and local projected over-density is separable in the two variables, suggesting that there are two dominant processes: one which depends on mass but not density (so-called ``mass-quenching'') and a second environment-related process which should be independent of stellar mass. The mass-quenching process is then the only one that depends on mass, and therefore is the one that determines the shape of the mass-function of the surviving star-forming galaxies and, via the continuity equation, the shape of the mass function of the resulting (mass-quenched) population of passive galaxies. The observed constancy of the shape of the mass function of star-forming galaxies over a wide redshift range up to $z \sim 2$ (or even higher) imposes a strong requirement on the form of mass-quenching (see P10 and below).

Subsequently, P12 showed that the environment-quenching in the overall population can be fully accounted by a satellite quenching process that applies only to satellite galaxies. The probability that a previously star-forming central galaxy is quenched when it becomes the satellite of another galaxy is about 50\%, independent of it's stellar mass. There are many possible suggestions for an environment-dependent quenching mechanism \citep[see e.g,][]{Gunn:1972p6714,McCarthy:2008p6864,Font:2008p6757}. 

The P10 prescription for mass-quenching can be written either as a quenching rate, i.e. the probability that a given star-forming galaxy will be quenched per unit time, or as a survival probability to reach a certain mass without being quenched. The probability $p_{\text{quench}}$ for a galaxy becoming quenched when increasing its stellar content by $\text{d}M_s$ is given by
\begin{equation} \label{eqn:quench}
	\text{d}p_{\text{quench}}=\mu{\text{d}M_s},
\end{equation}
for an infinitesimal $\text{d}M_s$. For a finite increase $\Delta M_s$, one gets
\begin{equation} \label{eqn:quench_finit}
	p_{\text{quench}}=1-\exp \left[-\mu{\Delta M_s}\right],
\end{equation}
The constant $\mu$ is required (see P10) to be $M^{\ast -1}$, where $M^{*} = 10^{10.68}$M$_{\odot}$ is the value of the characteristic stellar mass of the (single) Schechter stellar mass function of the blue star-forming population. Following P10, we take $\mu$ to be constant with time because $M^*$ is observed to be more-or-less constant. 

We will assume that the mass-quenching process acts on all galaxies, i.e. both centrals and satellites. This is motivated by the observational fact that $M^{\ast}$ is the same for central and satellite star-forming galaxies (P12). Because of the close coupling of stellar mass (and even BH mass) and halo mass for central galaxies, the action of a mass-quenching that is driven by stellar mass (as in the equation above) is hard to distinguish from one driven by halo mass for centrals, but again our point is that the outcome must be well represented by the empirical P10 quenching "laws".

For satellites, we apply an additional stochastic quenching process. When a central galaxy becomes the satellite of another galaxy because it's own halo merges with another more massive halo, the chance of it being (instantaneously) quenched is set to $p_{\text{sat}}=0.5$ . This additional quenching probability is only applied once to any particular galaxy when it first becomes a satellite. Because we do not, in the current paper, consider the radial distribution of galaxies within haloes \citep[e.g,][]{Prescott:2011p1365}, or try to compute the local over-density as in P12 or \cite{Kovac:2014p7312}, we do not consider the density-dependence of $p_{\text{sat}}$, instead adopting a mean value. This mean value of $p_{\text{sat}}=0.5$ is assumed to be constant with epoch, as shown in the zCOSMOS group catalogue to $z \sim 0.7$ \citep[][]{Knobel:2013p381,Kovac:2014p7312}.

To summarize, the quenching of galaxies in this model is accounted by just two constants, $\mu = 10^{-10.6}$M$_{\odot}^{-1}$ for mass-quenching and $p_{\text{sat}} = 0.5$ for satellite quenching.

\subsection{Dark matter structure formation}\label{sec:dm-process}

To describe the hierarchical structure formation process we take a simple model, far below the complexity of N-body simulation but aiming to account for most of the features of those simulations. The descriptions we apply have been incorporated in one or another way by many authors \citep[recently e.g. by ][]{Lu:2014p7573}. We use the dark matter merger tree generator from \cite{Parkinson:2008p31}, which is based on the excursion set theory \citep[e.g,][]{Press:1974p410, Epstein:1983p4176, Bond:1991p341, Lacey:1993p337} tuned to match the Millennium simulation \citep[][]{Springel:2005p331}. \cite{Parkinson:2008p31} showed that the tuned merger tree generator matches the overall halo mass function and the progenitor mass function for different halo masses very well back to redshift $z=4$. The merger tree generates its trees with a Monte Carlo method. Given a halo mass $M_h$ at redshift $z$ it generates the progenitors at $z+\Delta z$ for small time steps $\Delta z$ (backward process). In addition to a smoothed component growth there is a probability of having a binary split in the merger tree with a host and a satellite halo: 
\begin{equation}
	M_h \xrightarrow{\Delta z} M_{\text{host}} + M_{\text{sat}}+\Delta M_{\text{smoothed}},
\end{equation}
where $M_{\text{host}}$ is the most massive progenitor of $M_h$. The tree naturally divides the progenitors into a smooth component (all progenitors below a mass threshold $M_{\text{thresh}}$) and a merger component (growth due to accretion of mergers above $M_{\text{thresh}}$). We express the growth of a halo as
\begin{equation}
	\dot{M}_h=\dot{M}_{h,\text{smoothed}}+\dot{M}_{h,\text{merger}}.
\end{equation}

For the subhalo evolution we apply the formalism from \cite{BoylanKolchin:2008p1}. They used high resolution dark matter simulations with one host and one satellite halo to invert the dynamical friction time scale $t_{\text{df}}$ and provide a fitting formula for $t_{\text{df}}$ (Equations 5,6 in their paper with the further assumption that the last factor in their Equation 5 is equal to unity):
\begin{equation}
\label{eqn:t_df}
 \frac{t_{\text{df}}}{\tau_{\text{dyn}}}=0.216 \frac{(M_{\text{host}}/M_{\text{sat}})^{1.3}}{\ln(1+M_{\text{host}}/M_{\text{sat}})}e^{1.9\eta}.
\end{equation}
This formula depends on the host-to-satellite mass ratio $M_{\text{host}}/M_{\text{sat}}$ and
orbital circularity $\eta$. \cite{BoylanKolchin:2008p1} noted in their analysis that including the effect of baryonic bulges one gets an approximately 10\% shorter $t_{\text{df}}$. This fitting formula has been tested for $0.025\leq M_{\text{sat}}/M_{\text{host}}\leq 0.3$ and is applicable for $\eta \geq 0.2$.
Note that the dynamical time $\tau_{\text{dyn}}\approx 0.1H^{-1}$ with $H$ being the Hubble
parameter. The inverted dynamical friction time scale can be several times larger than the dynamical timescale $\tau_{\text{dyn}}$. From numerical simulations, \cite{Zentner:2005p290} have shown that the probability distribution of the orbital circularity $\eta$ of dark matter subhaloes can be approximated by
\begin{equation}
\label{eqn:zentner}
 P(\eta)\propto \eta^{1.2}\left(1-\eta\right)^{1.2}.
\end{equation}
For every merger event in our merger tree, we therefore draw $\eta$ from this distribution and thereby introduce some scatter in the dark matter structure formation process.

So far, we have an expression for the survival time $t_{\text{df}}$ of a subhalo. For the subhalo mass evolution $M_{\text{subhalo}}(t)$ we implement a step function following \cite{Yang:2012p243}
\begin{equation}
\label{eqn:m_subhalo_t}
 M_{\text{subhalo}}(t)= \begin{cases} M_{\text{sat}}\left(t=t_a\right) & t - t_a < t_{df}\\
  0 & t - t_a > t_{df},  \end{cases}
\end{equation}
where $t_a$ is the time of accretion.

In this paper, $M_h$ refers to the total halo mass. The halo mass associated with the central galaxy is then given by
\begin{equation} \label{eqn:central}
	M_{\text{central}}=M_h-\sum_i M_{\text{subhalo,i}},
\end{equation}
where the sum is over all surviving subhaloes above a certain mass threshold $M_{\text{thresh}}$. We thereby identify all substructure above $M_{\text{thresh}}$ and trace its evolution.

Our dark matter formalism clearly consists of some simplifications. The merger tree is tuned to a dark matter only simulation whereas our model contains baryonic matter too. One implicit simplification is that the baryonic matter component will not deviate from the behavior of dark matter. In other words, the gravitational forces from the dark matter are the dominant driver of baryonic structure formation and pressure terms are ignored. Likewise, there is no reverse effect from the baryons on the dark matter (see for example \cite{Borgani:2006p4146} for a more detailed description).

Also, it should be noted that the merger tree is tuned to a slightly different cosmology. However, the tuned parameters are dimensionless and as the excursion set approach is formulated for arbitrary power spectra, \cite{Parkinson:2008p31} argued that their merger tree can also be applied to different cosmologies. See also \cite{Jiang:2014p7547} for discussion on the accuracy. For the substructure evolution, we have applied a very simple description, especially for the time evolution of the substructure.  Despite these simplifications, our chosen description provides us with a good picture of what is going on in the dark matter structure formation process. It does not however contain the detailed and accurate descriptions that would be needed for doing precision cosmology.

To summarize, we introduced one arbitrary parameter $M_{\text{thresh}}$ in our structure formation model and take $t_{\text{df}}$ from \cite{BoylanKolchin:2008p1}.  The remaining parameters for the dark matter are taken from the standard cosmology.

\section{The model} \label{sec:combined}

In this section we describe how we combine all the ingredients given in Section \ref{sec:concepts}. In particular we describe in Section \ref{sec:infall_link} how we link the baryonic infall rate onto the regulator system to the dark matter structure formation process. In Section \ref{sec:gal_gal_merging} we describe and discuss what happens in a galaxy-galaxy merging event in our model framework. In Section \ref{sec:breack_down} we describe how the regulator at very low stellar masses can be described. The procedure to predict the cosmic abundances of galaxies and their properties is described in Section \ref{sec:sample_creation}. Finally we emphasize in Section \ref{sec:no_freedom} how our model differs from others parametric approaches.

\subsection{Link between baryonic and dark matter infall rate} \label{sec:infall_link}

To consistently integrate our regulator and quenching models into the dark matter framework, some further assumptions have to be made. First, only dark matter haloes and subhaloes above $M_{\text{thresh}}$ in Section \ref{sec:dm-process} will contain a regulator system. In other words, we ignore star-formation in haloes that are so small that we considered their infall as part of the smooth dark matter inflow. This is because they will be mostly gaseous. We set $M_{\text{thresh}}=1.4\cdot10^{9}M_{\odot}$. This is somewhat arbitrary, but is consistent with photo-ionisation heating suppressing cooling and star formation below a certain halo mass $M_{\gamma}$. $M_{\gamma}\sim 10^8M_{\odot}$ during reionisation to $M_{\gamma}\sim \text{few}\cdot 10^9M_{\odot}$ \citep[][]{Gnedin:2000p1139,Okamoto:2008p1151}. For a more realistic model aiming to make predictions of low mass galaxies back to the epoch of reionisation, one would need to account for a change in the mass threshold. We explore the effect of changing $M_{\text{thresh}}$ in the Appendix and show that it is small for the galaxy mass scales of interest.

In order to trace the gaseous baryons through the build-up of haloes, the following simple scheme was used. We will later refer to this model as Model A.

\begin{itemize}

\item 
First, all gaseous baryons in a given halo are associated at all times with one of the regulator systems (i.e. ``galaxies'') within that halo {\it except} for those baryons which have been processed through a regulator and ejected from the galaxy through the wind described by $\lambda$ in Section \ref{sec:regulator}. These ejected baryons are assumed to be ``lost'' (we will revisit this assumption later in the Paper) and are no longer tracked. But, apart from this, all gaseous baryons are found within the reservoirs of the regulator systems.

\item 
Second, when two haloes merge, the baryons that are at that time within each of the regulator systems in the haloes stay within those regulators, unless the (sub)halo subsequently decays and is disrupted (see below).

\item 
Finally, smooth accretion of gas onto haloes, i.e. the baryonic inflow associated with the merging of haloes below $M_{\text{thresh}}$, is split between the sub-haloes as follows:

\end{itemize}
\begin{equation} \label{eqn:infall}
	\Phi_{b,i}=f_b\dot{M}_{h,\text{smoothed}}\cdot\frac{M_{\text{subhalo,i}}}{M_h}.
\end{equation}

This scheme ensures that every baryon which has not flown into some regulator in the past, will be assigned to a regulator when coming into a halo above $M_{\text{thresh}}$. It also ensures that when a regulator becomes a satellite, its infall rate and thus its SFR will not dramatically change, as observed (see P12). We note that when a galaxy is quenched, the gas inflow associated with this quenched galaxy will not be redirected to other active regulators. In our discussion later in the paper, we will introduce a different assignment of the in-flowing gas and see how this will change our predictions, in what we will call Model B. Realizing that the gas inflow description is crucial to many observables we will then introduce a further Model C, which provides far more freedom in assigning gas to regulator systems.

As noted above, L13 considered only a single regulator in a given halo and introduced $f_{\text{gal}}$ as the fraction of inflowing baryons that penetrate down and enter the regulator system at the center of the halo. L13 concluded that $f_{\text{gal}} \sim 0.5$ was required to reproduce the stellar to dark mass ratio of typical galaxies. By associating all gaseous baryons to regulator systems, we are effectively setting $f_{\text{gal}}$ to unity (i.e. eliminating this parameter) in Model A and B in the present paper. However, because we now include multiple regulators (associated with the subhaloes) in a given halo and a two component growth (mergers and smoothed accretion), the net effect for the central regulator will be similar because only a fraction ($\dot{M}_{\text{h,smooth}}/\dot{M}_{\text{h}}$) of the halo growth is associated with gas accretion and it will only receive a fraction ($M_{\text{central}}/M_{\text{h}}$) of the incoming gas. In other words, we would now understand that the adoption of the lower $f_{\text{gal}} \sim 0.5$ in L13 was simply accounted for the two component growth of a halo, which was neglected in their treatment of regulator systems.

\subsection{Subhalo disruption / galaxy-galaxy merging} \label{sec:gal_gal_merging}

We now turn to what happens when a subhalo decays according to Equation (\ref{eqn:m_subhalo_t}) and specifically what happens to the gas and stars within the regulator associated with that sub-halo. The two extreme cases would be adding all the stars and gas to the central galaxy or distributing them into the inter-cluster medium, which for the gas would involve re-distributing the gas amongst the surviving regulators according to Equation \ref{eqn:infall}. Reality is likely in between these extremes. For concreteness and convenience, we set the fraction of stars and gas which are given to the central galaxy $f_{\text{merge}}=0.5$ but show in Appendix \ref{ap:merging} that the output of the model is insensitive to this parameter. When the gas and stellar component from two different regulator systems is merged in this way, the new state of the regulator will likely not be in equilibrium with the gas infall rate. Galaxy-galaxy merging can thus lead to some scatter in the regulator properties. As discussed in L13 and illustrated in their Figure 3, the regulators adopt quickly to the new conditions and rapidly settle to the new equilibrium state.

\subsection{Break down of the regulator description at low $M_s$} \label{sec:breack_down}

In L13, the parameters of the regulator (Equation \ref{eqn:epsilon} and \ref{eqn:lambda} in this paper) were tuned to match the metallicities of galaxies with stellar masses above $10^8M_{\odot}$ and this parameterization must break down at lower stellar masses: not least, the mass-loading cannot increase without limit, simply on energetic grounds. We however need to include such low mass galaxies in our model so as to have larger galaxies later on. We therefore introduce a maximal outflow load. We set $\lambda_{\text{max}}=50$. This value is far off the regime where L13 tuned their parameters and therefore will not affect the validity of the tuning in L13. It also does not significantly affect the output of the model for galaxies above $M_s=10^{8}M_{\odot}$, the mass range of primary interest. Further discussion of this parameter can be found in Appendix \ref{ap:outflow}.

\subsection{Implementation} \label{sec:sample_creation}

It will have been clear that the input galaxy data going into the model was derived independently of the number of galaxies, i.e. specifically it was the mean mass-SFR-metallicity relation (L13), the shape of the star-forming mass function parameterized by $M^*$ (P10), and the red fractions of satellites (P12). A primary output of the model will be the expected number density of galaxies.

We therefore need to create a representative sample of the Universe. Merger trees derived from N-body simulations are sampled according to the halo mass function and therefore produce far more low mass halo trees than for high mass haloes. As we want to achieve the same statistical power over a wide range in halo mass, we want to equally sample the halo masses and weight their abundances in a second step. The merger tree generator provides such a possibility. The procedure is as follows: We sample 10'000 haloes at redshift $z=0$, chosen randomly from a flat distribution in logarithmic halo mass, from $7.1 \cdot 10^{9}M_{\odot}$ up to $1.4 \cdot 10^{15}M_{\odot}$. We then weight their abundance according to the halo mass function of \cite{Sheth:1999p463} at $z=0$. By construction, the weighted abundance of our haloes is then in perfect agreement with the input halo mass function at $z=0$. We then let these haloes run backwards in cosmic time by applying the merger tree description. We stop when our resolution limit $M_{\text{thresh}}$ is reached or at $z=15$. At that point we identify our regulator systems, put in some initial stellar and gas mass and solve the differential equations for every single tree component. In parallel we apply the subhalo evolution model in the forward process. We thereby keep track of every satellite halo with its own regulator system. The model is not sensitive to the initial state of the regulators, as described in Appendix \ref{ap:init_cond}).

Clearly, this description has no spatial resolution, either within galaxies, within haloes or to follow the large scale distribution of haloes. The last of these would be relatively easy to implement and this will be the subject of a future paper. The other two would take us deeper into details, which we wish to avoid.

\subsection{A model without re-adjusting the parameters} \label{sec:no_freedom}

In Table \ref{tab:parameters} all of the parameters of our model are listed with a short description and reference to the input data on which they are based. These are mainly taken from the three papers P10, P12 and L13, and from cosmology and computational simplifications in the dark matter sector. The effects of the three additional parameters that we have introduced in this paper, i.e. $M_{\text{thresh}}$, $f_{\text{merge}}$ and $\lambda_{\text{max}}$, are investigated in the Appendices \ref{ap:merging}, \ref{ap:outflow} and \ref{ap:m_thresh}. We conclude there that any reasonable variation within these parameters do not invalidate our conclusions. In essence, these parameters are introduced for practical reasons to make the model operable and the output does not depend very much on their precise values.

Within our chosen gas inflow description we therefore have virtually no freedom in changing our predictions: The model either matches observations or produces a discrepancy from which we may hope to learn. The goal is therefore not at first to produce a model that fits all available data, nor to observationally determine parameters. Rather, and in the spirit of the previous papers (P10,P12 and L13) we aim instead to provide insights into how well the ideas presented in those papers perform in the global context of a dark matter hierarchy, and to see where we encounter limitations.

\begin{deluxetable*}{lllrr}
\tabletypesize{\scriptsize}
\tablecaption{This table lists all our model parameters (for Model A and B), its values and to what set of data they are tuned to.}
\tablewidth{0pt}
\tablehead{
Symbol & Description & Fixed to: & Units  & Value 
}
\cutinhead{Regulator Parameters (externally derived)}
$\epsilon_{10}$      & Efficiency normalization to     & Metallicity data \tablenotemark{a} & $Gy^{-1}$ & 0.33   \\
          & $10^{10}M_{\odot}$ stellar mass       \\
$b$       & Power law of efficiency     & Metallicity data \tablenotemark{a} & - & 0.3      \\
	      & as a function of stellar mass          \\
$\lambda_{10}$      & Outflow load normalization to     & Metallicity data \tablenotemark{a}  & - & 0.3   \\
          & $10^{10}M_{\odot}$ stellar mass       \\
$a$       & Power law of outflow load     & Metallicity data \tablenotemark{a} & - & -0.8      \\
	      & as a function of stellar mass          \\
\cutinhead{Quenching parameters (externally derived)}
$M^*$     & Mass-quenching parameter $\mu^{-1}$     & Exponential cutoff   & $M_{\odot}$ & $10^{10.68}$\\
	& & of main sequence \tablenotemark{b}\\
$p_{\text{sat}}$ & satellite quenching probability & Elevated ref fraction of satellites \tablenotemark{c} & - & 0.5\\
\cutinhead{Additional practical parameters in the combined model}
$f_{\text{merge}}$ & merging fraction of gas and stars & Parameter with no significant  & - & 0.5 \\
		  & of disrupted subhaloes & effect on our conclusions \tablenotemark{e} \\
$\lambda_{\text{max}}$  & Maximum outflow load  & Upper bound provided by  & - & 50 \\
					& of regulator &  regulator action in tuning range \tablenotemark{f} \\
$M_{\text{thresh}}$ & Threshold in halo mass  & Photo-ionisation model \tablenotemark{g} & $M_{\odot}$ & $1.4\cdot10^{9}$ \\
					& for having a regulator & \\
\cutinhead{Cosmological Parameters (externally derived)}
$h$  & dimensionless Hubble parameter  & CMB \tablenotemark{d} &  -  & 0.7 \\
$\Omega_b$ & Baryonic density  & CMB \tablenotemark{d} & - & 0.45 \\
$\Omega_m$ & Matter density & CMB \tablenotemark{d} & - & 0.3 \\
$\Omega_{\lambda}$ & Dark Energy density & CMB \tablenotemark{d} & - & 0.7 \\
$\sigma_8$ & Power spectrum normalization & CMB \tablenotemark{d} & - & 0.8 \\
$n_s$ & spectral index & CMB \tablenotemark{d} & - & 1.0 \\
\cutinhead{Additional simplification descriptions of the Dark Matter sector}
$t_{\text{df}}$ & dynamical friction time scale & Dark Matter N-body simulation \tablenotemark{h} & - & Eq \ref{eqn:t_df}\\
$\eta$ & orbital circularity & Dark Matter N-body simulation \tablenotemark{i} & - & Eq \ref{eqn:zentner}\\
\tablenotetext{a}{Data from \cite{Mannucci:2010p318} fitted by L13}
\tablenotetext{b}{Data and model fit by \cite{Peng:2010p132}}
\tablenotetext{c}{From \cite{Peng:2012p1015}, \cite{Kovac:2014p7312} and \cite{Knobel:2013p381}}
\tablenotetext{d}{From WMAP seven-year data \citep[][]{Komatsu:2011p966}}
\tablenotetext{e}{Further discussion in Appendix \ref{ap:merging}}
\tablenotetext{f}{Further discussion in Appendix \ref{ap:outflow}}
\tablenotetext{g}{Model by \cite{Gnedin:2000p1139} and \cite{Okamoto:2008p1151}, further discussion in Appendix \ref{ap:m_thresh}}
\tablenotetext{h}{Relation from \cite{BoylanKolchin:2008p1}}
\tablenotetext{i}{Relation from \cite{Zentner:2005p290}}
\label{tab:parameters}
\end{deluxetable*}

\section{Results}

\label{sec:results}

In Section \ref{sec:concepts}, we reviewed the different and independent inputs that were then combined in Section \ref{sec:combined} to produce a single model of star-formation and quenching in galaxies within a dark matter hierarchical framework. In this section, we compare the output of the default model A with both observations directly and with the outputs of other phenomenological approaches to galaxy evolution, most notably that of \cite{Behroozi:2013p7464}. 

As discussed above, we will not vary any pre-adjusted parameter in our model beyond the three parameters introduced to allow the model to be computed (the values of which do not much affect the outcome) and so we can examine these comparisons one at a time. Throughout this section, we refer always to the same output sample generated with the parameters given in Table \ref{tab:parameters} with the inflow description of Equation \ref{eqn:infall}, referred as our fiducial Model A.

It should be noted that the observational data used to determine these parameters were (a) gas metallicity data (as in L13) from \cite{Mannucci:2010p318} SDSS, specifically the Z($M_s$,SFR)-relation, (b) the red fraction of satellites (as in P12 from \cite{Abazajian:2009p5798} SDSS DR7) and (c) the value of M* of star-forming galaxies (as in P10 also from SDSS). Any predictions of these particular quantities must therefore match observations, by construction, but predictions of all other quantities are bona fide and can be meaningfully compared with other data.

Comparison of these predictions with other data will enable us to draw several interesting conclusions. Some of the successes of these ``predictions" will mirror conclusions that were already drawn in the original papers on which our new model is based, e.g. the discussions of mass functions and red fractions in P10 and P12, and the link between sSFR and specific accretion rate in L13. For these, it is reassuring to see them holding up in the context of a more realistic treatment of the haloes, including substructure and merging etc. None of the predictions based on the {\it population} of dark matter haloes could be made before, since they were not treated in the earlier works. These include the normalization of the mass functions and the computation of the star-formation rate density. We can also predict the scatter in various relations comming from different halo assembly histories. 

Finally, we will make explicit comparisons with the output from the orthogonal phenomenological approach of \cite{Behroozi:2013p7464}. The \cite{Behroozi:2013p7464} approach is anchored in the dark matter hierarchy and derives a very general description of the effect of baryonic processes within these haloes. In that work, a general $M_s/M_{h}$ relation is assumed. The epoch dependent form of this is then derived by simultaneously applying statistical tools such as abundance matching of the mass functions at different redshifts, coupled with comparison of the consequent information on star-formation with a variety of observational data, including the sSFR$(M_s,t)$ and the global star-formation rate density SFRD. Our own approach is in a sense orthogonal to this as it is based on a prior determination of the purely baryonic phenomenology which is then imported into the dark matter structure. Despite the quite different approaches, and the obvious limitations of each of them, we will find that a very similar picture emerges.

\subsection{Stellar Mass dependence of the Main Sequence sSFR at the present-day} \label{sec:SFR:main_sequence}

\begin{figure}
  \centering
  \includegraphics[angle=0, width=80mm]{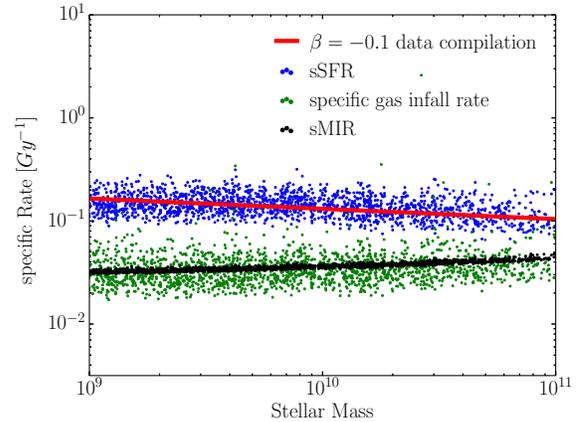}
  \caption{Prediction of the sSFR-mass relation at redshift $z = 0$ for blue central galaxies in Model A (blue). Dots correspond to individual galaxies in the model. The red line corresponds to the relation given in Equation \ref{eqn:sSFR_z} from a data compilation in slope and normalization (for citations see text). The green dots correspond to the specific gas infall rate of the same galaxies. Black dots denote the specific mass growth rate of the entire halo (sMIR) on timescales shorter than the last major merger event. The mean in the specific gas infall rate is the same as the sMIR. The sSFR is elevated by more than a factor of two. The scatter in the sSFR follows the scatter in the gas infall rate.}
\label{fig:SFR_stars_scatter}
\end{figure}

We first plot in Figure \ref{fig:SFR_stars_scatter} the specific star formation rate (sSFR) of all blue (i.e. star-forming) central galaxies of the output sample at $z=0$ as a function of their stellar masses. The model successfully recovers the tight correlation between sSFR and mass which is known as the Main Sequence \citep[e.g,][]{Brinchmann:2004p5336, Noeske:2007p706} and an almost constant sSFR with a scatter about this relation of about 0.2 dex.

For comparison with data we over-plot an sSFR$(M_s,z)$ relation of the form
\begin{equation}
	\text{sSFR}\propto M_s^\beta.
\end{equation}
Observational estimates of $\beta$ range between $-0.4 < \beta < 0.0$ at stellar masses above $10^9M_{\odot}$, with most estimates $\beta \sim -0.1$ \citep[e.g,][]{Brinchmann:2004p5336, Noeske:2007p706, Elbaz:2007p824, Daddi:2007p837, Pannella:2009p1509, Stark:2013p7595, Peng:2010p132}. In Figure \ref{fig:SFR_stars_scatter} the red line illustrates the data compilation in the form
\begin{equation}\label{eqn:sSFR_z}
	\text{sSFR}(M_s,z)=0.12 \left(\frac{M_s}{10^{10.5}M_{\odot}} \right)^\beta (1+z)^3    \text{       (at } z<2)
\end{equation}
with $\beta=-0.1$ evaluated at $z=0$ (see L13 and references therein). The observed scatter amongst real galaxies is about 0.3 dex once outliers with much higher sSFR are excluded \citep[see e.g,][]{Rodighiero:2011p1606, Sargent:2012p7578}. These latter are associated with star-bursts, probably induced by mergers.

The mean sSFR$(M_s)$ at $z = 0$ is clearly well reproduced by the model. As noted in L13 and discussed earlier in this paper, a key feature of the kind of gas regulation considered in this paper is that it sets the sSFR close to the specific mass accretion rate of the system, independent of the values of the parameters $\epsilon$ and $\lambda$ controlling the regulator. There is a modest ``boost" to the sSFR if an individual regulator system is increasingly efficient at producing stars as time passes (as would be expected if the efficiency increases with mass). This boost at $z = 0$ is expected to be of order 0.3 dex for typical galaxies. It increases to lower masses, potentially reversing the slope of the sSFR$(M_s)$ relation relative to that of the specific accretion rate, defined as sMIR$= \dot{M}_h/M_h$. L13 took the approximation for the sMIR provided by \cite{Neistein:2008p335}. Despite our model using a more complex description for the baryonic infall rate $\Phi_b$, we would expect to have the same underlying link between the sMIR and sSFR. The good agreement with the mean $z = 0$ sSFR$(M_s)$ relation in the current model which contains a wide variety of individual haloes is therefore reassuring but not unexpected given the discussion in L13 (see their Fig 9).
 
The scatter in sSFR$(M_s)$ in our model is caused by the different halo formation histories, i.e. by the variation in the gas inflow rate caused by variations in the merger tree (green dots in Figure \ref{fig:SFR_stars_scatter}), and by the effects of galaxy-galaxy merging (see Section \ref{sec:gal_gal_merging}). Our model does not include any further stochastic time-variation in the gas infall $\Phi_b$ such as might be caused by other baryonic processes, and also neglects any stochastic scatter in the baryonic processes controlling star-formation within the galaxy regulator systems. Both of these could further increase the scatter (in our model there is almost no scatter occurring in the SFR-$\Phi_b$ relation). Our predicted scatter can therefore be interpreted as a lower bound in the expected sSFR$(M_s)$ scatter. The fact that it is already 2/3 of the observed scatter suggests that these two further contributors to the scatter (stochastic infall variability and variation in the regulator) can contribute only of order 0.2 dex in normal Main Sequence galaxies. 

\subsection{Epoch dependence of the Main Sequence sSFR and the star-formation rate density} \label{sec:sSFR_results}

\begin{figure}
  \centering
  \includegraphics[angle=0, width=80mm]{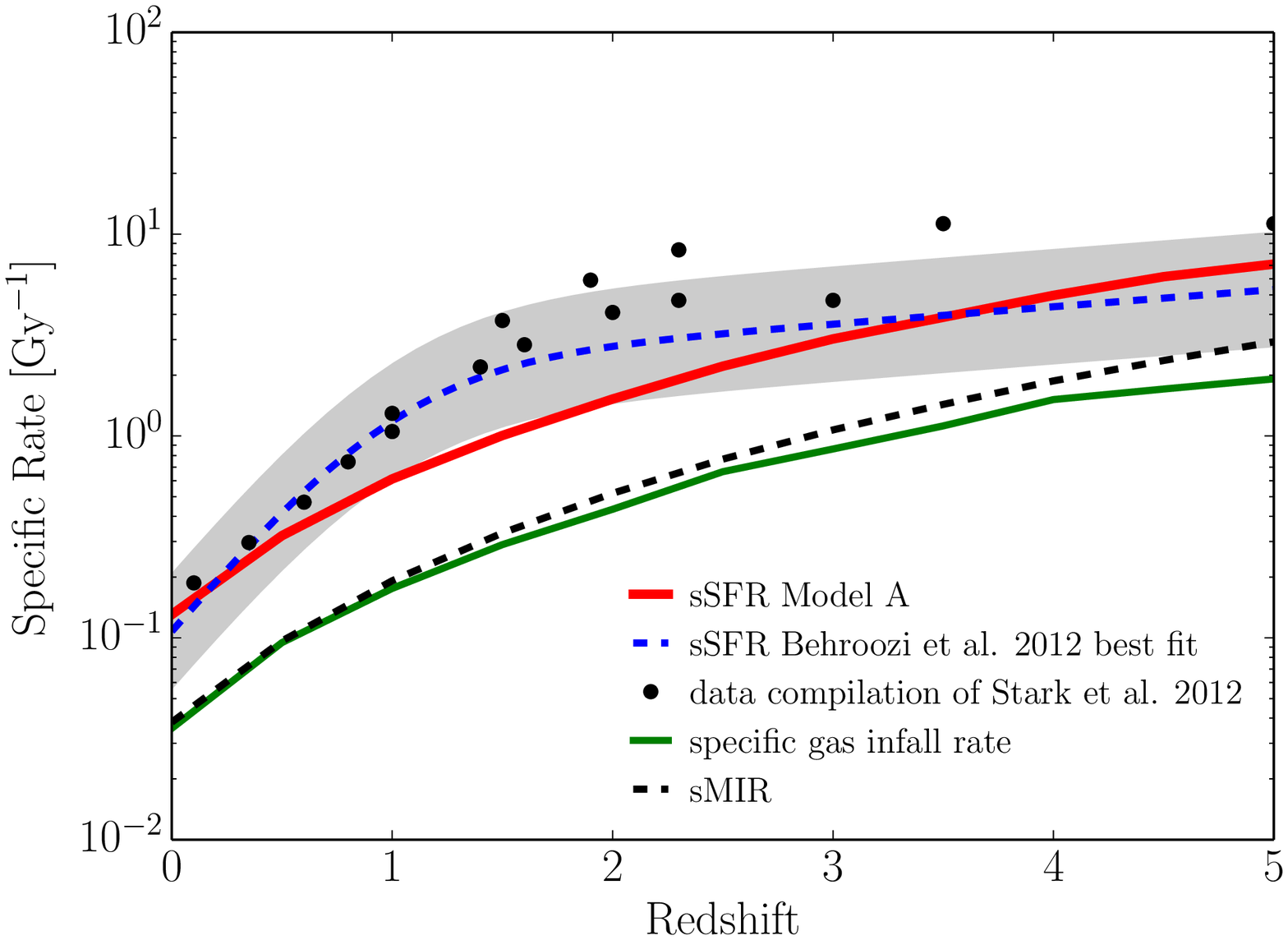}
  \caption{Prediction of the mean sSFR for blue galaxies within a stellar mass range of $10^{10}M_{\odot}$-$10^{10.5}M_{\odot}$ of Model A as a function of redshift (red curve). These are compared with data points (in black without errorbars) from \cite{Stark:2013p7595} and a model based on a data compilation of \cite{Behroozi:2013p7464} adjusted to our definition of SFR. The gray region reflects the 1-$\sigma$ scatter between different measurements in the literature given by \cite{Behroozi:2013p7464}. The specific gas infall rate of the same galaxy sample of our model is over-plotted in green. The sSFR follows this quantity with an offset (boost) as discussed in L13. Furthermore the specific mass increase rate of the halo (sMIR) is over-plotted.}
\label{fig:sSFR_sMIR}
\end{figure}

In Figure \ref{fig:sSFR_sMIR}, we show the evolution in the sSFR for galaxies in the mass range $10^{10}M_{\odot}$-$10^{10.5}M_{\odot}$ back to $z = 5$, compared with data from \cite{Stark:2013p7595} and a highly parameterized model of \cite{Behroozi:2013p7464} adopted to our definition of sSFR=SFR/$M_s$. We also show for comparison the mean sMIR and the specific gas infall rate. As expected the sSFR tracks the increase in sMIR with redshift. While this broadly matches the data, the rise with redshift is not steep enough. As a result, the observed sSFR at $z \sim 2$ is about a factor of two higher than predicted from the model.

This is a common problem encountered in galaxy evolution models \citep[e.g,][and others]{Weinmann:2012p1729, Dave:2011p1971} and is also present in the simple analysis of L13 that used an average halo growth rate from \cite{Neistein:2008p335}. Adjustment of the prediction would require a substantial modification of the accretion rate of baryons onto the regulator systems, i.e. breaking the link between the baryonic accretion rate onto the galaxy and the specific growth rate of the dark matter halo) or a rather dramatic adjustment of the efficiency with which inflowing gas is converted to stars (i.e. the $f_{\text star}$ parameter of L13) so as to increase the boost factor associated with temporal changes in this quantity (see L13). We will return to this discrepancy in models B+C but note here that it is not inconceivable that some of the offset of 0.3 dex could reflect observational difficulties in determining stellar mass and star formation rates at high redshifts.

Our model naturally produces a deviation of the baryonic increase rate to the dark matter growth rate at very high redshifts as the dynamical friction time scale cannot catch up the halo growth rate resulting in far more substructure surrounding the central at high redshifts. More substructure means within our model that less baryonic infall will be assigned to the central as described in Equation (\ref{eqn:infall}).

\begin{figure}
  \centering
  \includegraphics[angle=0, width=80mm]{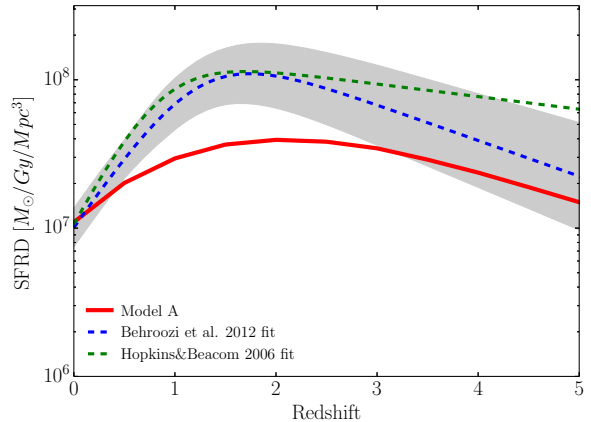}
  \caption{The Star Formation Rate Density (SFRD) from our model sample as a function of redshift (red line). The SFRD is the integrated SFR over all galaxies at a certain cosmic time, normalized to unit volume. The blue dashed line is the best fit model of \cite{Behroozi:2013p7464} and the gray region is the 1-$\sigma$ inter-publication scatter noted by them. The green dashed line is the best fit of the data compilation of \cite{Hopkins:2006p1430}. Our model predicts the right normalization at $z=0$ and the drop in the SFRD at late times. Our model does not reproduce the boost in SFRD at $z=2$ in its full strength.}
  \label{fig:SFR_tot}
\end{figure}

In Figure \ref{fig:SFR_tot}, the overall star formation rate density (SFRD) is plotted over the whole range of cosmic time compared with data from the compilation by \cite{Hopkins:2006p1430} and the phenomenological model by \cite{Behroozi:2013p7464}. The gray region is the 1-$\sigma$ inter-publication scatter noted by \cite{Behroozi:2013p7464}.

The broad features of the evolving SFRD of the Universe are reproduced and our predicted value at $z=0$ matches well the observational data of the nearby Universe. We again see a tension in the model that the SFRD is too low at $z = 2$. The size of the discrepancy is roughly the same as for the sSFR($z$) evolution. We return to this below.

\subsection{The evolution of the gas fraction in galaxies}\label{sec:gas_star_evolution}

\begin{figure}
  \centering
  \includegraphics[angle=0, width=80mm]{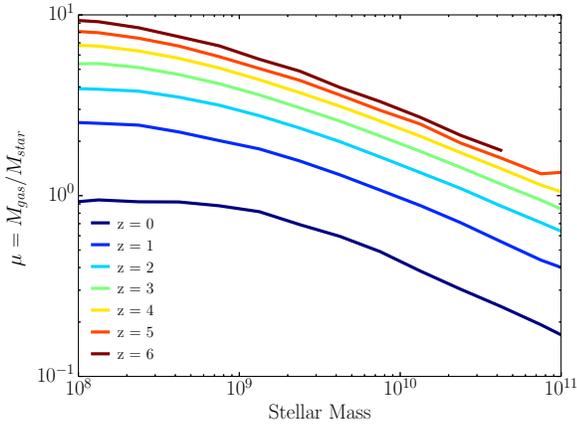}
  \caption{The gas-to-star ratio is plotted as a function of stellar mass at different redshifts for Model A. With our default regulator parameters, this ratio is increased at $z = 4$ by about a factor of six relative to locally.}
  \label{fig:gas_stellar}
\end{figure}

In Figure \ref{fig:gas_stellar}, we plot the gas-to-star ratio $\mu=M_{\text{gas}}/M_s$ as a function of stellar mass for different redshifts. We get about a factor of six higher gas-to-star ratio at $z\sim 4$ compared to $z=0$. 
From the definition of the regulator quantities in L13, the gas ratio is simply given by the ratio of the sSFR and the star-formation efficiency $\epsilon$

\begin{equation} \label{eqn:mu}
	\frac{M_{\text{gas}}}{M_s} = \frac{sSFR}{\epsilon}.
\end{equation}
So the increase in the gas ratio is a direct result of the fact that the halo growth rate and thus the sSFR increases faster with redshift than the dynamical time of the galaxy which was assumed to set the redshift evolution of $\epsilon$. Lowering the gas fraction in high redshift galaxies can be done in two different ways: One either lets the efficiency $\epsilon$ increase faster with redshift or as a higher power of the gas mass within the regulator. These have similar effect because of the higher gas fractions at high redshift.

In our model A, the gas infall rate $\Phi_b$ drops faster with cosmic time than the star formation efficiency $\epsilon$ and therefore galaxies become less gas-rich at later cosmic times (a similar argument was drawn in \cite{Dave:2011p81}). This behavior is in qualitative agreement with observations \citep[e.g,][]{Tacconi:2010p1222,Geach:2011p1249}.

\subsection{Stellar Mass Function (SMF)} \label{sec:smf}

The galaxy stellar mass function (SMF) is a well measured quantity at low redshifts \citep[e.g,][]{Baldry:2008p1191, Pozzetti:2010p2191, Peng:2010p132, Baldry:2012p2212}. Our model provides predictions for the overall SMF and also for the population split into blue and red galaxies (i.e. star-forming and quiescent) and into centrals and satellites. As noted above, the model is constructed to reproduce the characteristic Schechter cutoff of the blue population at $M^* \sim 10^{10.68}$M$_{\odot}$ and for this to be constant with time, but we have not introduced any other parameter that is based on e.g. the faint end slope of the blue and red population, or the red fraction at $M^*$). The mass quenching law of P10 can directly predict the relative faint end slopes of the blue and red population, but the absolute slope $\alpha_{s,blue}$ of the blue population had to be assumed. The red fraction at M* also follows from the input $\alpha_{s,blue}$.

\begin{figure}
  \centering
  \includegraphics[angle=0, width=80mm]{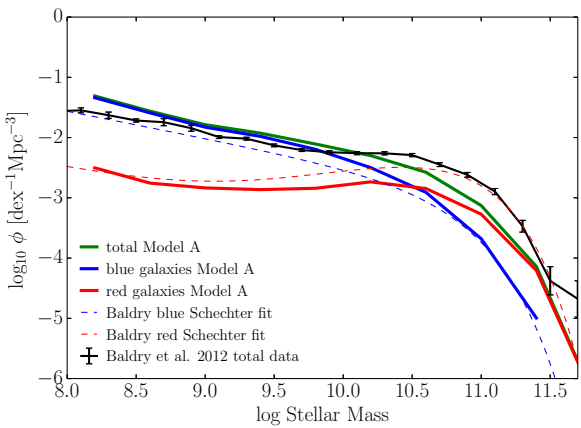}
  \caption{The Stellar Mass Function at $z=0$ is plotted for Model A. The green line is the overall SMF from our Model A. The blue curves are for the blue population, the red curves for the red population (including centrals and satellites). The output is compared to the data of \cite{Baldry:2012p2212}. Dashed lines corresponds to Schechter fits to the blue and red population in their paper.
}
\label{fig:Behroozi}
\end{figure}

\begin{figure*}
  \centering
  \includegraphics[angle=0, width=160mm]{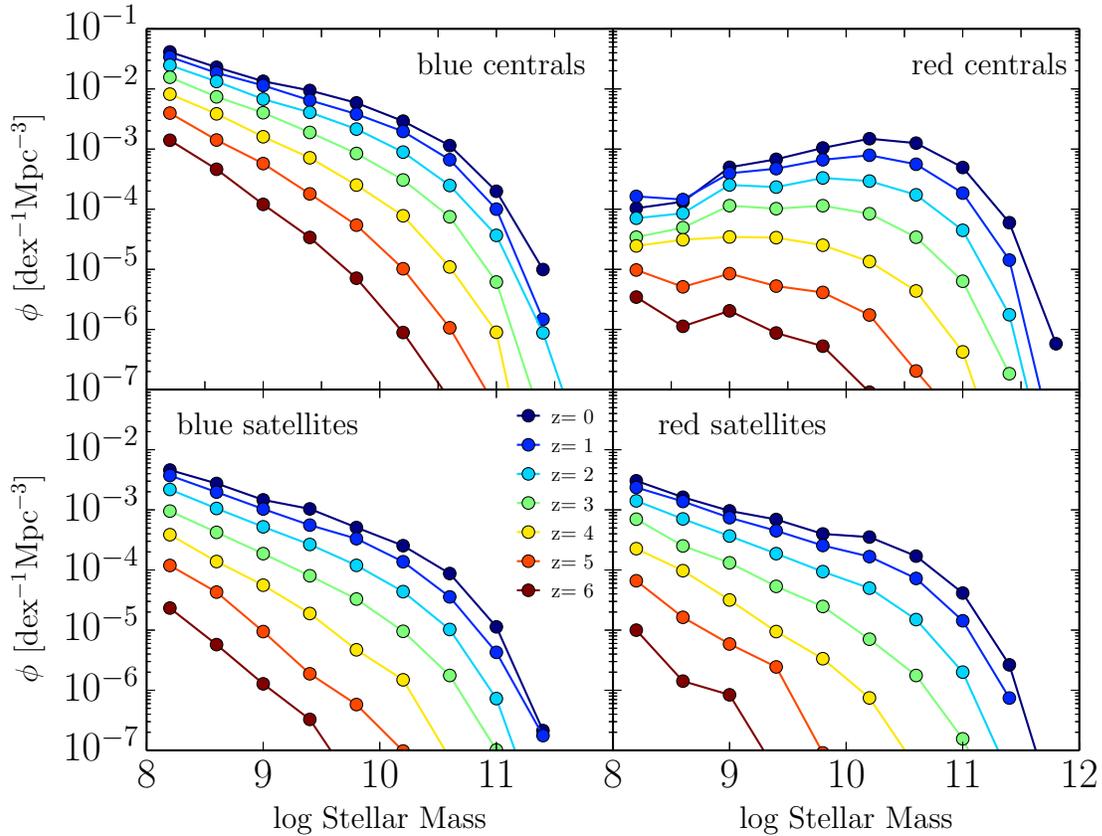}
  \caption{The SMF for the blue and red population, centrals and satellites are plotted for redshifts $6\leq z\leq 0$ for Model A. Our model predicts a nearly constant fraction of blue and red population, centrals and satellites back to at least $z=4$. The Schechter parameters for these SMF's are given in Appendix \ref{ap:tables}.}
  \label{fig:smf_evolution}
\end{figure*}

In Figure \ref{fig:Behroozi} the model prediction for the blue, red and total population at $z = 0$ is plotted, while in Figure \ref{fig:smf_evolution}, we present our results for the evolution of the SMF's for different galaxy types (split into red and blue and into central and satellite) over cosmic time. The Schechter parameters for these SMF's of the red and blue centrals and satellites are given in Appendix \ref{ap:tables}. The red satellite population can be better described by a double Schechter function. As shown in P12, this is due to superposition of mass- and satellite-quenching (more about the fits in Appendix \ref{ap:tables}).

The model successfully reproduces the correct faint end slope of the mass function. This is a reflection of the link between the slope of the mass-metallicity relation and the faint-end slope $\alpha$ of the mass-function (see L13 for discussion). The relations between the Schechter parameters ($M^*$ and $\alpha$) of the different populations in Figure \ref{fig:smf_evolution} are also as observed. The universality of $M^*$ (all populations have very similar $M^*$) and the change in faint end slope $\Delta\alpha \sim 1.0$ between blue and red centrals, are also successfully reproduced. These follow from the forms of the quenching laws derived in P10 and P12.

Less trivial is the overall normalization of the SMF of the different populations.  The $\phi^*$ describes the normalization at $M^*$ in the Schechter function fits.  The SMF is the convolution of the stellar-to-halo mass relation (SHMR), including its scatter, with the underlining halo mass function.  We note that the underlying halo mass function is Press-Schechter like and not Schechter like. If we do not apply the mass quenching description, the SMF prediction would look Press-Schechter like and would have a rapidly evolving characteristic mass. At very high redshift, where the galaxy population could not build up a significant fraction of galaxies with stellar masses above $M^*$, we predict a Press-Schechter like SMF. In our model we see that the transition from a Press-Schechter to a ``vertically evolving'' Schechter-like SMF happens between $z=6$ and $z=4$ (from Figure \ref{fig:smf_evolution}). It is the moment when the stellar mass functin breaks away the halo mass function. \cite{Lilly:2013p7458} referred to this as the Phase 1 to Phase 2 transition. We can also clearly see that the satellite population grows more rapidly with cosmic time than the one of the centrals in Fig \ref{fig:smf_evolution}, also indicated by the Schechter fits in the Appendix \ref{ap:tables}. This means that the special role of the quenching of satellite galaxies becomes more and more important with cosmic time. The satellite-quenching leads to the double-Schechter component in the SMF of the red population. The differential rate of quenching of the two populations and the fact that the quenched satellites dominate at lower masses leads to the appearance of ``down-sizing'' , i.e. a more gradual buildup of the stellar mass-function at lower masses.

The biggest problem with the mass functions is a surprising one. Although the shape of the mass function of passive galaxies is right, their overall number density is too low. This also produces a weaker bump in the ``double" Schechter function that is caused by the superposition of the red and blue SMF (which have different faint end slopes $\alpha$). This is surprising because one of the great successes of the P10/P12 quenching formalism was to explain, via the continuity equation, the ratio of these two components, which is given simply as $(1+\alpha)^{-1}$ where $\alpha$ is the faint end slope of the star-forming mass function. For $\alpha \sim -1.4$ this would predict a ratio of about 2.5, close to what is observed, whereas our model predicts more like 1.5 . But we clearly note that with $\alpha \sim -1.5$ (our Schechter fit) the ratio goes already down to about 2.0 .
We will return to discuss this interesting question further in Section \ref{sec:discussion}.

\subsection{Star formation rate history in different mass haloes and the evolution of the star-formation rate density}

\begin{figure}
  \centering
  \includegraphics[angle=0, width=80mm]{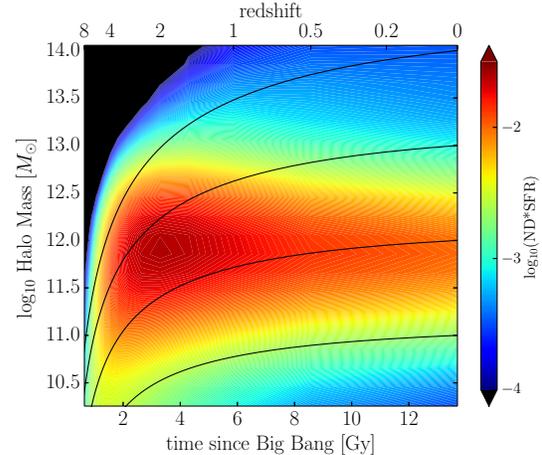}
  \caption{The Star Formation Rate history as a function of cosmic time and halo mass for Model A. The plot includes SFR from centrals and satellites. Black lines indicate an average growth history of different haloes. Units of the color scale are chosen to be ND*SFR dlog$_{10}(M_h)=M_{\odot}yr^{-1} Mpc^{-3}$dlog$_{10}(M_h)$.}
  \label{fig:sfr_halo}
\end{figure}

We now turn to comparisons with the phenomenological model of \cite{Behroozi:2013p7464}. In Figure \ref{fig:sfr_halo}, we show our prediction for the SFR in haloes (including centrals and satellites) as a function of cosmic time and halo mass. This may be compared with the similar Figure 4 from \cite{Behroozi:2013p7474} which was derived from their completely different but similarly phenomenological approach.

\cite{Behroozi:2013p7464} concluded that most stars were formed around $z=2$ in haloes of about $10^{12}M_{\odot}$. This is a natural output of our model as the regulator is highly inefficient in producing stars at low stellar masses and (mass-)quenching is most effective above $M_s=M^*$, which corresponds to about $10^{12}M_{\odot}$ in halo mass. 

The fact that these two orthogonal approaches produce broadly the same phenomenological picture is very reassuring. It furthermore emphasizes the operational difficulty of distinguishing, for central galaxies, whether the dark matter mass or the (baryonic) stellar mass is driving the variable efficiency with which haloes convert baryons into stars, simply because these two quantities are tightly linked. 

\subsection{Stellar-to-halo mass relation (SHMR)}

One of the central properties of galaxies is the stellar-to-halo mass relation (SHMR), both for centrals and for satellite galaxies. The SHMR represents the overall efficiency with which haloes convert baryons into stars. This quantity has been extensively studied using abundance matching and other statistical techniques such as halo occupation distributions, which are based on the conviction that the SHMR should be well-behaved. Observations using weak-lensing can be used to directly test these, generally with success \citep[e.g,][]{Leauthaud:2012p159}.

\begin{figure*}
  \centering
  \includegraphics[angle=0, width=160mm]{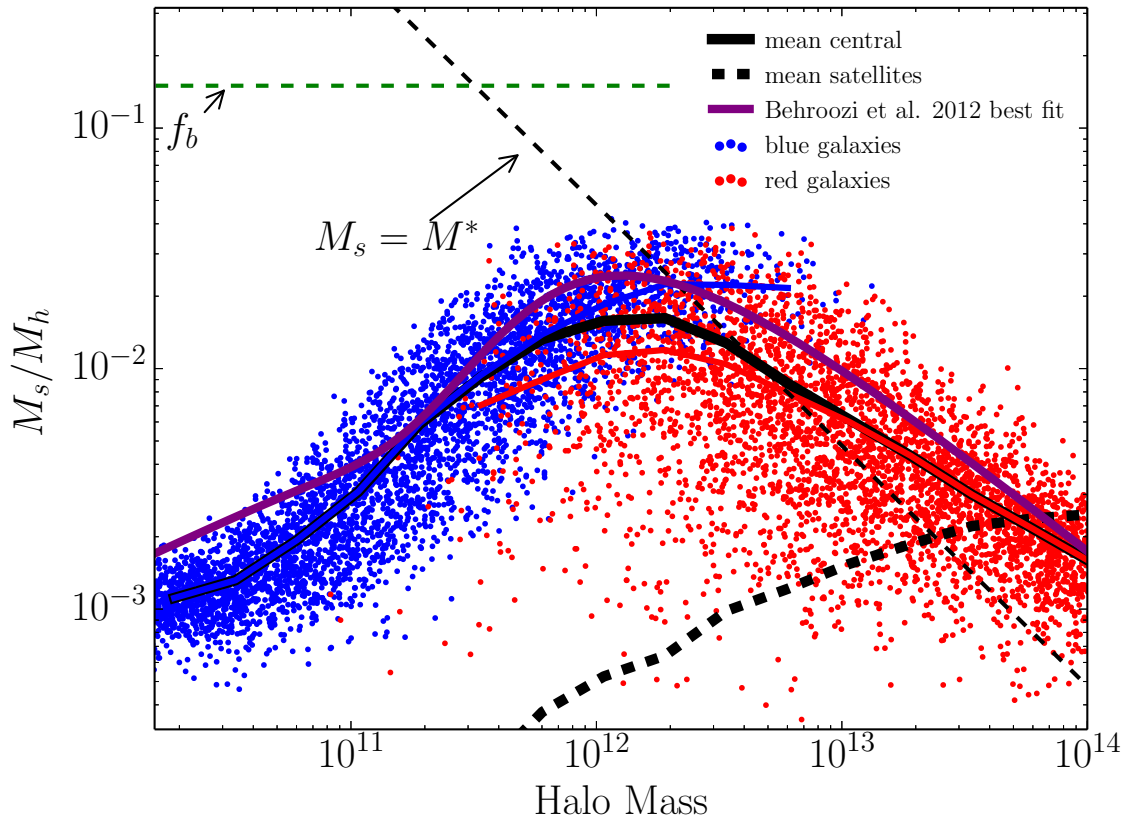}
  \caption{The SHMR at $z=0$ of Model A is plotted as a function of the total halo mass $M_h$ for the set of central galaxies, separated into red and blue. The blue (red) continuous line is the mean value of the blue (red) population in our model and the black line is the mean SHMR of the overall sample for centrals (i.e. a suitably weighted average of the red and blue lines). The thick dotted black line is the contribution of satellites to the SHMR while the green thin dotted line indicates the cosmic baryonic fraction. The global turn-over of the star formation efficiency can fully be accounted by quenching galaxies around $M^*$, corresponding to about $10^{12}M_{\odot}$ in halo mass. The agreement with the abundance matching reconstruction of \cite{Behroozi:2013p7464} is quite impressive, although there is a systematic reduction in $M_s/M_{h}$ above $M_h \sim 10^{11.5}M_{\odot}$ which may be traced to the saturation of the efficiency with which the regulator in L13 converts baryons to stars that is in turn linked to the flattening of the $Z(M_s)$ mass-metallicity relation.}
  \label{fig:sh_scatter}
\end{figure*}

The SHMR for our output sample at the present epoch is plotted in Figure \ref{fig:sh_scatter} and compared with the zero-redshift relation from \cite{Behroozi:2013p7464}. As would be expected, the increase in the $M_s/M_{h}$ ratio at low masses simply reflects the increasing efficiency of converting baryons to stars (i.e. $f_{\text{star}}$ in L13) in more massive regulators, while the turn-over and subsequent decline is due to the mass-quenching of galaxies which becomes progressively more important at masses around and above $M^*$, corresponding to about $10^{12}M_{\odot}$ in halo mass. 

The 1-$\sigma$ scatter in the SHMR of the blue population in the model is about 0.21 dex. This comes mostly from the different halo assembly histories (e.g, the time when the last major merger happened). The scatter in the red population is larger and is about 0.36 dex. This ultimately reflects the quite broad range in stellar (or halo) mass over which central galaxies have been mass-quenched and the continued growth of haloes after the star-formation has been quenched. 

Red galaxies have systematically lower $M_s/M_{h}$ than blue ones at a given $M_h$ because their stellar masses are frozen at quenching (apart from mass growth due to merging) while their dark matter haloes continue to grow. They may scatter down the $M_s \sim M^*$ locus. This scatter explain the observation \citep[e.g,][]{Woo:2013p2295} that at a given stellar mass, red galaxies are found in higher mass haloes (e.g. with more satellites). As the overall population of central galaxies changes from predominantly blue at low halo masses to predominantly red at higher halo masses the mean SHMR shifts from that of the blue galaxies to that of the red. The overall scatter is expected to be 0.32 dex at the peak but deviates from being a log-normal distribution in stellar mass.

Overall, the agreement between the output of our model and the reconstruction from \cite{Behroozi:2013p7464} is very good. Our curves for the overall population are slightly lower around the peak, by up to about 0.2 dex at halo masses above $10^{11.5}M_{\odot}$ and this can be traced to the saturation of $f_{\text{star}}$ in L13, which itself was driven by the saturation in the adopted $Z(M_s)$ mass metallicity relation. We will return to this point below and show that it is closely linked to the issue of the deficit of quenched galaxies noted in Section \ref{sec:smf}.

\begin{figure*}
  \centering
  \includegraphics[angle=0, width=160mm]{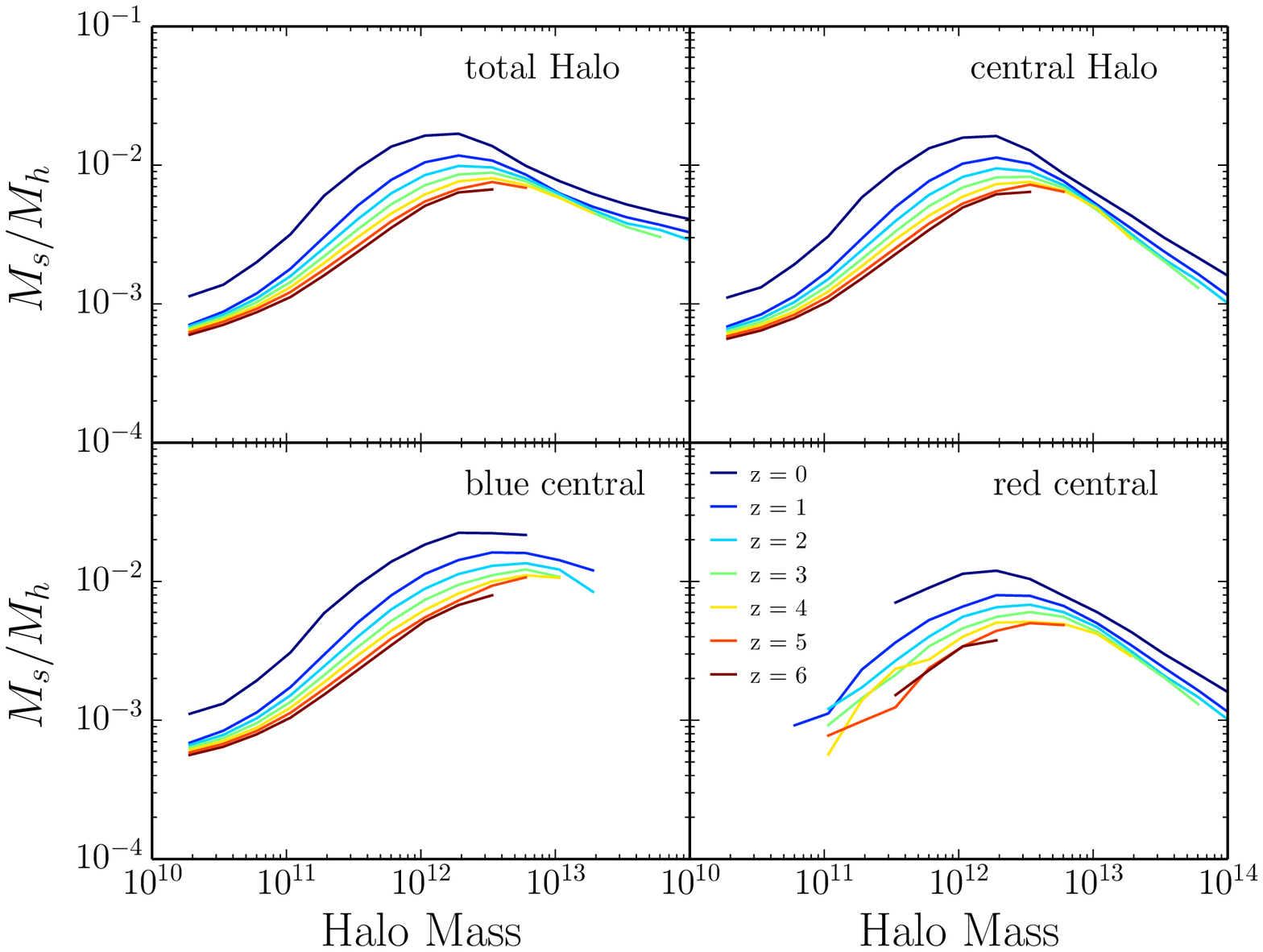}
  \caption{The average SHMR of Model A for different galaxy types as a function of total halo mass. Upper left: Considering all galaxies within a halo and all types of galaxies. Upper right: Considering only the central galaxy (red and blue). Lower left: Considering only blue central galaxies. Lower right: Considering only red central galaxies. The scatter in this relation can be deduced from Figure \ref{fig:sh_scatter}. The SHMR is predicted to be an increasing function in cosmic time (i.e. decreasing with increasing redshift) because regulators at high redshift are more gas rich.}
  \label{fig:stellar_halo}
\end{figure*}

Our model has a slight redshift evolution in the SHMR (see Figure \ref{fig:stellar_halo}). Within our model, this is due to the fact, that regulators (i.e. galaxies) at higher redshifts contain proportionally more gas and thus less stellar mass as discussed in Section \ref{sec:gas_star_evolution}. But the general behavior remains at all redshifts the same. At very low halo masses, the stellar content remains dominated by the maximum outflow load $\lambda_{\text{max}}$ and the saturation feature occurs at every redshift at roughly the same halo mass. The nominal drop in the SHMR at $z=4$ is about a factor of two.

\section{Discussion}

\label{sec:discussion}

In Section \ref{sec:results} we recovered a number of encouraging agreements of various predictions compared to the literature, both in terms of observational data and in terms of the independent and orthogonal phenomenological model of \cite{Behroozi:2013p7464}. In particular there is no reason for the total number density of galaxies to come out right. The models and the parameters taken from the previous papers (P10, P12, L13) did not have any information about the abundance of dark matter haloes nor were designed to match the number density of galaxies in the universe. This is a remarkable success of our model. The model is simple but still reproduces a wide range of non-trivial results. In this section, we will have a closer look at those areas where our model produces discrepancies that may give clues as to where additional features could be added, or which may highlight more fundamental tensions.

First we will have a look at the specific star formation rate evolution and note how we can, in principle, achieve a better agreement with the data compilation of \cite{Stark:2013p7595} and \cite{Behroozi:2013p7464} in Section \ref{sec:sSFR_evolution}. But then, relating the sSFR evolution to the SFRD evolution, we argue that we can not easily bring these two observations in agreement with each other, independent of our model assumptions (Section \ref{sec:SFRD_link}). We then turn our attention to the missing red galaxies. We discuss how this is linked to the form of the SHMR in Section \ref{sec:sh-relation} and we discuss its relation to the saturation feature of the L13 regulator model. In Section \ref{sec:new_infall} we propose two other ways of assigning the gas in-flow to the galaxies within the halo and see that we get a further improvement in matching the SMF, sSFR and SFRD history, our Model B and C. In Section \ref{sec:coincidence} we discuss a very specific feature of our models and finally in Section \ref{sec:abundance_matching} we relate our results to abundance matching methods.

\subsection{Modification to match the sSFR at z=2} \label{sec:sSFR_evolution}

\begin{figure}
  \centering
  \includegraphics[angle=0, width=80mm]{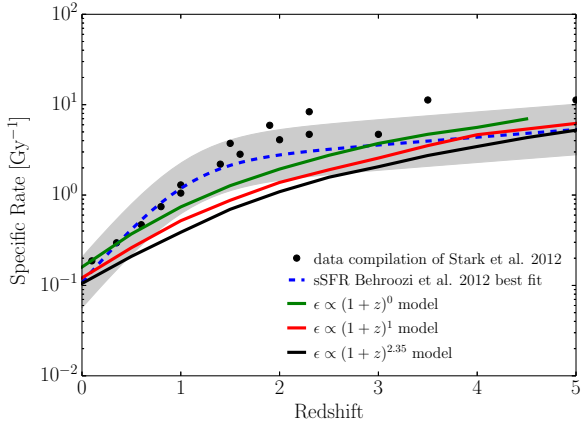}
  \caption{The sSFR history for three different variants of the redshift evolution of the star-formation efficiency $\epsilon$ with $c=0,1,2.35$ (as defined in Equation \ref{eqn:epsilon_z}) are plotted. Stronger evolution (higher c-values) lead to a lower sSFR at all times.}
  \label{fig:compare_sSFR}
\end{figure}

In Section \ref{sec:sSFR_results} and in Figure \ref{fig:sSFR_sMIR} we noted a deviation of the sSFR evolution at $z=2$ between our predictions and the data compilation of { \cite{Stark:2013p7595} and \cite{Behroozi:2013p7464} }. It might be thought that one possible way of modifying our model to try to get a better match is to change the star-formation efficiency, $\epsilon$, at high redshift. Detailed discussion about the link between star formation and gas reservoir has been made by several authors \citep[recently e.g,][]{Feldmann:2013p5526}.
However, because the link between sSFR and the sMIR (specific mass accretion rate of the system) is independent of $\epsilon$ and $\lambda$ (see L13, and thus also of $f_{\text{star}}$), modification of $\epsilon(z)$ changes the sSFR only through the ``boost'' effect on sMIR that is associated with a change in $f_{\text{star}}$ with time and so the effect of this change should be quite weak. It turns out that a higher $\epsilon$ at high redshift leads to a drop in the offset of sSFR compared with the sMIR. To explore this, we modify the parameterization of $\epsilon$ to:

\begin{equation} \label{eqn:epsilon_z}
	\epsilon(z)\propto \left(1+z \right)^c
\end{equation}

with $c$ being the additional model parameter. In our default Model A (also Models B and C below), the efficiency scales as the Hubble rate. In Figure \ref{fig:compare_sSFR} we plot three different models with $c=0,1,2.35$, i.e. assuming no redshift evolution, one coming close to the fiducial model and one in which the efficiency scales as the sMIR according to \cite{Neistein:2008p335}. We note that at fixed redshift, the efficiency is parameterized as a function of $M_s$. This parameterization is fitted at $z=0$ and might not provide a direct link to the physical process that actually sets the efficiency. 

We clearly see that {\it lowering} the star-formation efficiency at higher redshifts actually {\it boosts} the sSFR. This is because it lowers $f_{\text{star}}$ at high redshifts and therefore increases the boost term in Equation 36 of L13. On the other hand if the efficiency increases with redshift as fast as the specific infall rate, we reduce the sSFR. In both cases, the effect of the change in the sSFR is spread out over a wide range of redshifts (because of the smooth evolution in $\epsilon$) and we cannot get a peak at one particular redshift, or drastically change the overall slope.

An alternative approach is to decouple the specific accretion rate onto the regulator systems from the specific growth rate of the surrounding dark matter haloes. A redshift dependent cold gas accretion efficiency \citep[e.g,][]{Bouche:2010p311} could do this, or some other scheme to limit the baryonic accretion onto the regulators. In Section \ref{sec:new_infall} we will explore some modifications by introducing Models B and C.

\subsection{The link between sSFR and SFRD} \label{sec:SFRD_link}

\begin{figure}
  \centering
  \includegraphics[angle=0, width=80mm]{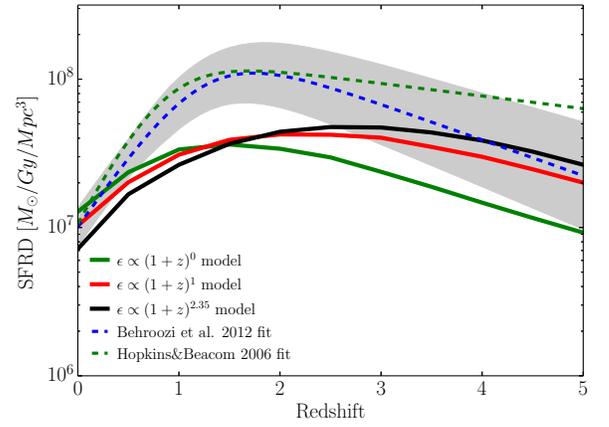}
  \caption{The SFRD history for the three different variants of $\epsilon(z)$ with $c=0,1,2.35$ (defined in Equation \ref{eqn:epsilon_z}). Stronger evolution (higher c-values) leads to an enhancement of the SFRD at early times and a stronger decrease at later times.}
  \label{fig:compare_SFRD}
\end{figure}

Staying with the same expansion in our model as in Section \ref{sec:sSFR_evolution} we turn our attention to the star formation rate density (SFRD). We plot in Figure \ref{fig:compare_SFRD} the SFRD history for the same three models as for Figure \ref{fig:compare_sSFR}. The figure shows that lowering the efficiency at high redshift shifts star formation to later times. The redshift dependence of the efficiency $\epsilon$ does not have a significant influence on the outcome at $z=0$.   It has a slight effect of where the stellar mass is formed. As the model has a smoothed evolution in $\epsilon$, significant diviations in the sSFR history from our default model can not be made.

\subsection{Matching the red fraction at $M^*$} \label{sec:sh-relation}

As mentioned in \ref{sec:smf}, our model under-predicts the abundance of red galaxies around $M^*$. In other words, the relative fraction of red to blue galaxies is too low. 

The number density of red galaxies around $M^*$ is directly related to the number of dark matter haloes between $M_h(M^*)$ and infinity. As the halo mass function is a very steeply decreasing function of halo mass, the number of red galaxies around $M^*$ is very sensitive to the halo mass $M_h(M^*)$ that corresponds to the quenching mass $M^*$. 

However, simply changing the parameter $M^*$ (i.e. $\mu^{-1}$) will have a severe impact on the blue population that we match very well. Boosting the SHMR (e.g, by just letting more gas flow in the regulator) is also not satisfactory. By doing so, we will boost the number density of blue and red satellites by the same amount. We would be able to get the needed number density in the red population around $M^*$ (as we lowering the halo mass corresponding to $M^*$) but at the same time we would end up with to many blue galaxies at the same stellar mass range. The question is: How can one change the red fraction without either changing the number density of the blue population or $M^*$? The fraction between blue and red galaxies around $M^*$ is dependent on how fast galaxies are approaching $M^*$. We have to elevate the sSFR at $M^*$ or in terms of the SHMR, the power law parameter for the Main sequence $\gamma$ defined as
\begin{equation}
	M_s\propto M_h^\gamma
\end{equation}
has to be steeper around $M^*$ than our model prediction. 

Our model produces a flattening of the SHMR around $M^*$ (see Figure \ref{fig:sh_scatter} for $z=0$ and Figure \ref{fig:stellar_halo} for the redshift evolution). This is an intrinsic feature of the regulator model and independent of quenching. The overall fraction of baryons in stars cannot exceed the cosmic fraction, and indeed can only asymptotically approach this. In fact, because of the ``loss'' of outflowing gas in this first Model A, it will saturate at an even lower value. The regulator $f_{\text{star}}$ saturates when the gas within the halo is nearly used up.

We note that our model, even without any quenching mechanism, therefore has a saturation feature coming from the regulator because $f_{\text{star}}$ is limited to some value. Our model predicts just at the stellar mass when quenching happens a flattening of $\gamma$ due to the saturation. In contrast, we get a better match to the red population when abandoning the saturation feature or invoking an even steeper $\gamma$ at $M^*$. This might provide a hidden link between the quenching process and the running out of gas of the galaxy. We return to this below.

\subsection{Changing gas in-flow description} \label{sec:new_infall}

One of the weaknesses of our models is that we do not trace the out-flowing gas. The need for gas reincorporation in a cosmological context was initially analysed in \citep[][]{Benson:2003p5972, DeLucia:2004p6956}. Other recent works include \citep[][]{Oppenheimer:2008p6820,Oppenheimer:2010p6858,Henriques:2013p2980}. In our simple model, we don't allow the expelled gas to get back into the same regulator or transfer it to another regulator sitting in the same dark matter halo. Letting some or all of this gas back into the regulator system will change the output of our model significantly. We note that at stellar masses around $M^*$, about 1/2 $M^*$ of gas has been ejected earlier in the history of each galaxy. There is only a slight dependence of this on the adopted value of the parameter $\lambda_{\text{max}}$. From our discussion in Section \ref{sec:sh-relation}, the saturation feature leads to a mismatch of the red population. To delay the saturation of our regulator to higher stellar masses above $M^*$, we might just put some of the ejected gas back into the regulator at the time when saturation occurs. This process can in principle be accomplished by setting an appropriate recycling time (of order several dynamical times). Such a behavior can consistently be applied to our model. The only worry is that this new type of metal-enriched inflow will significantly change the metallicity-fitted parameters inferred in L13 and used in our combined model. This might indicate that the metallicity modeling might be unrealistic.

Some gain in the direction can be achieved by simply modifying how gas is assigned to the regulators. In combining the different models of Section \ref{sec:concepts} we have a freedom in assigning the gas in-flow to the different galaxies (central or satellites). So far in our Model A we have assigned the gas according to the weights of the (sub)haloes (Equation \ref{eqn:infall}) with the weight of the central given in Equation \ref{eqn:central}. The substructure fraction is increasing with halo mass and therefore the second term in Equation \ref{eqn:central} assigns a smaller proportion of the infallen gas to the central galaxy as it grows in stellar mass. This can also contribute to the flattening of the SHMR.

The Model A assumed no domination of the central galaxy over its satellites at all. The other extreme would be the central galaxy dominates completely and gets all the gas in-flow and the satellites do not get any gas infall at all. Our Model B which we present here is identical to our Model A except that Equation \ref{eqn:infall} is changed so that all of the incoming gas is given to the central galaxy:

\begin{equation} \label{eqn:new_infall}
	\Phi_{b,i}= \begin{cases} f_b\dot{M}_{h,\text{smoothed}} & \text{central}\\
	 0 & \text{satellite}.  \end{cases}
\end{equation}

\begin{figure} 
  \centering
  \includegraphics[angle=0, width=80mm]{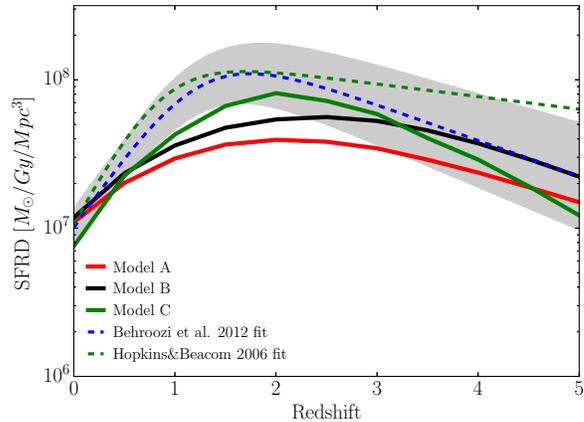}
  \caption{The same plot as Figure \ref{fig:SFR_tot}. The SFRD for Model A of Section \ref{sec:results} (red), Model B (black) and Model C (green) are compared with data compilations. Model B achieves some boost around $z=2$ compared to Model A, but only Model C achieves the required amount of boost.}
  \label{fig:SFRD_tuned}
\end{figure}

The result in terms of the SFRD is plotted in Figure \ref{fig:SFRD_tuned}. We clearly see an additional boost in the SFRD around z=2 or even at higher redshift. This brings the model closer to what is required by the data. The reason for the difference between the two proposed models is that at high redshift the halo merger rate is very high compared to the subhalo decay rate. This leads to more substructure within a halo at high redshift. In our Model A this leads to less gas in-flow onto the central galaxy, which is avoided in Model B. Furthermore the gas infallen onto the central galaxy is turned into stars more efficiently than in (lower mass) satellites. But despite this improvement, the Model B still under-predicts the SFRD at $z=2$. 

\begin{figure} 
  \centering
  \includegraphics[angle=0, width=80mm]{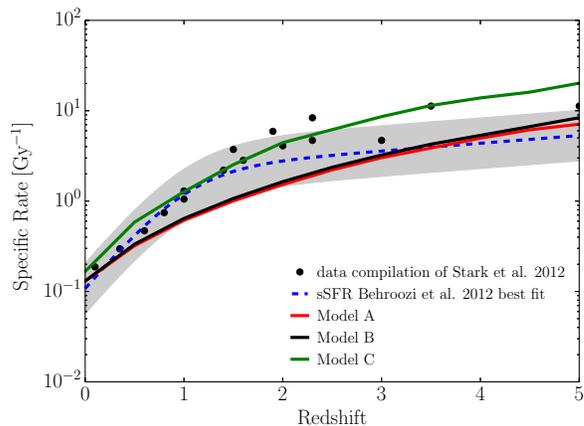}
  \caption{The same plot as Figure \ref{fig:sSFR_sMIR}. The sSFR for Model A of Section \ref{sec:results} (red), Model B (black) and Model C (green) are compared with data compilations. Model B is very similar around $z=2$ compared to Model A but only Model C achieves the required boost.}
  \label{fig:sSFR_tuned}
\end{figure}

In terms of the sSFR history we do not get any change in the predictions form Model A to Model B, as presented in Figure \ref{fig:sSFR_tuned}. To match the sSFR history, we have to change the model further.

\begin{figure}
  \centering
  \includegraphics[angle=0, width=80mm]{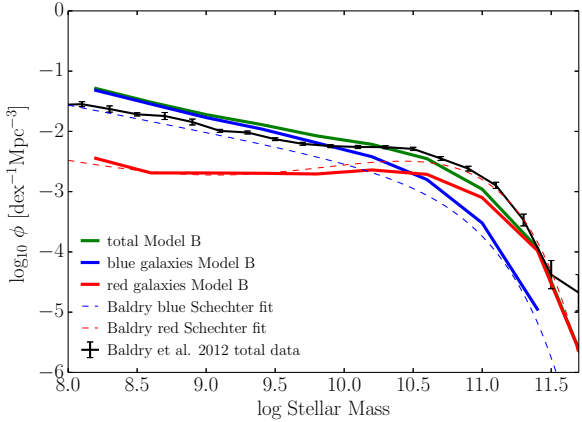}
  \caption{The Stellar Mass Function at $z=0$ is plotted for Model B. The green line is the overall SMF. The blue curves are for the blue population, the red curves for the red population (including centrals and satellites). The output is compared to the data of \cite{Baldry:2012p2212}. Dashed lines corresponds to Schechter fits to the blue and red population in their paper.}
  \label{fig:SMF_model_B}
\end{figure}

\begin{figure}
  \centering
  \includegraphics[angle=0, width=80mm]{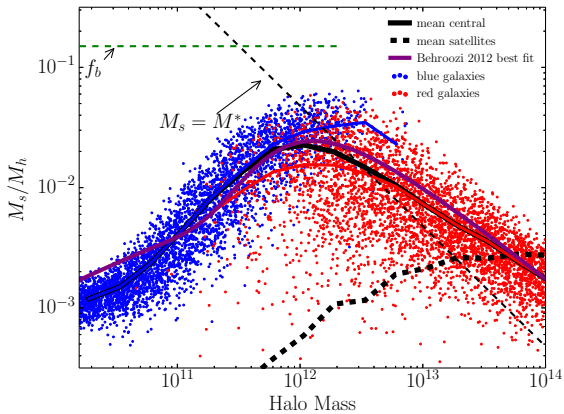}
  \caption{The SHMR at $z=0$ of Model B is plotted as a function of the total halo mass $M_h$ for the set of central galaxies, separated into red and blue (same as Figure \ref{fig:sh_scatter} for Model A). The blue (red) continuous line is the mean value of the blue (red) population in our model and the black line is the mean SHMR of the overall sample for centrals (i.e. a suitably weighted average of the red and blue lines). The thick dotted black line is the contribution of satellites to the SHMR while the green thin dotted line indicates the cosmic baryonic fraction.}
  \label{fig:SHMR_model_B}
\end{figure}

Looking at the SMF at $z=0$ predicted by our Model B in Figure \ref{fig:SMF_model_B} we can also partially improve matching the red fraction around $M^*$. A discrepancy remains, however, coming from the regulator description as discussed in Section \ref{sec:sh-relation}. 
The SHMR of Model B (Figure \ref{fig:SHMR_model_B}) for central galaxies is similar to Model A and also comes close to the Model of \cite{Behroozi:2013p7464}.

Out of this discussion, we see the importance of how one assigns the gas in-flow to the different galaxies within a halo. But we want to emphasize that no complicated description (e.g recycling of outflown gas, decoupling of baryonic inflow and dark matter growth, ...) is needed to achieve the level of agreement that is already presented in Model A and B. In terms of the quenching "laws", they are instead to be purely descriptive. These laws would likely be more complicated if they were formulated in terms of physical mechanisms which are still unclear.

Having said that, the red fraction problem and the sSFR and SFRD at $z=2$ still do not match perfectly. Our next approach is the one of an "effective SAM". From our discussion above, we concluded that the gas inflow description is crucial in perturbing our model and, doing it in the right way, matching the observables. For our Model C we introduce a redshift and halo mass dependent gas inflow. We change equation \ref{eqn:infall} to the form:

\begin{equation} \label{eqn:new_infall_2}
	\Phi_{b,i}= \begin{cases} f_b\dot{M}_{h,\text{smoothed}}\cdot f_a(a)\cdot f_M(M_{\text{i}}) & \text{central}\\
	 0 & \text{satellite},  \end{cases}
\end{equation}
with
\begin{equation}\label{eqn:f_m}
	f_M(M_h)=1+30\cdot \left(\frac{M_h}{10^{12}M_{\odot}}\right)^{2.5}
\end{equation}
and
\begin{equation} \label{eqn:f_gal_a}
	f_{\text{z}}(z)= \begin{cases} -1.25\cdot (1+z)^{-1} + 1.4 & z < 2\\
	0.25 + 6.75\cdot(1+z)^{-3} & z >= 2.  \end{cases}
\end{equation} 
The functions $f_M$ and $f_z$ are arbitrary and designed to have four desirable features:
\begin{enumerate}
	\item $f_z$ is a decreasing function between $z=2$ and $z=0$ accounting for the steep decline in the SFRD.
	\item $f_z$ is a rapidly increasing function approaching $z=2$ accounting for the boost in the sSFR around $z=0$.
	\item $f_M$ has an additional term such that there is significantly more gas inflow onto massive galaxies around $M_h(M^*)$ to counter-act the saturation feature of the regulator.
	\item $f_z\cdot f_M$ is normalized such that the baryonic mass within the regulator never exceeds the cosmic baryonic fraction of the universe.
\end{enumerate}

\begin{figure}
  \centering
  \includegraphics[angle=0, width=80mm]{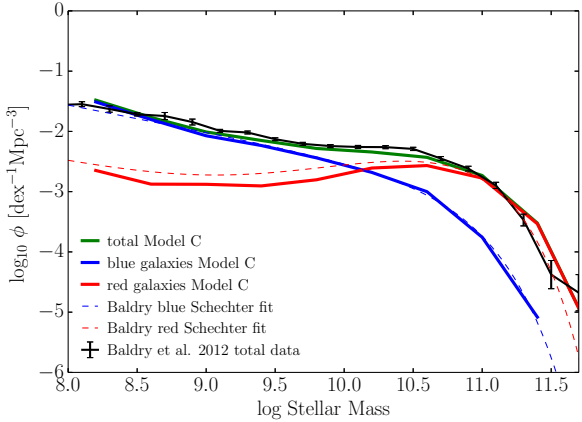}
  \caption{The Stellar Mass Function at $z=0$ is plotted for Model C. The green line is the overall SMF. The blue curves are for the blue population, the red curves for the red population (including centrals and satellites). The output is compared to the data of \cite{Baldry:2012p2212}. Dashed lines corresponds to Schechter fits to the blue and red population in their paper.}
  \label{fig:SMF_model_C}
\end{figure}

\begin{figure}
  \centering
  \includegraphics[angle=0, width=80mm]{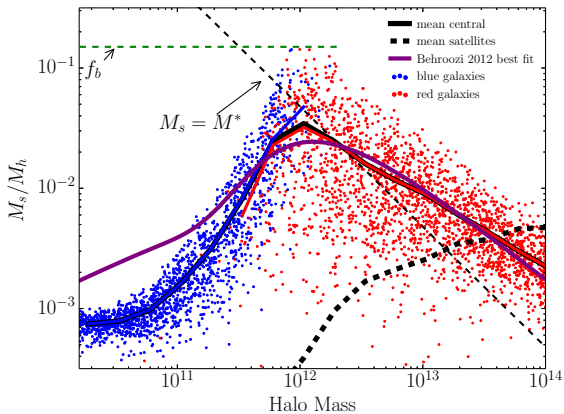}
  \caption{The SHMR at $z=0$ of Model C is plotted as a function of the total halo mass $M_h$ for the set of central galaxies, separated into red and blue (same as Figure \ref{fig:sh_scatter} for Model A). The blue (red) continuous line is the mean value of the blue (red) population in our model and the black line is the mean SHMR of the overall sample for centrals (i.e. a suitably weighted average of the red and blue lines). The thick dotted black line is the contribution of satellites to the SHMR while the green thin dotted line indicates the cosmic baryonic fraction. This model best matches the abundance of red and blue galaxies. Quenching occurs just when the blue galaxies are approaching the cosmic baryonic limit.}

  \label{fig:SHMR_model_C}
\end{figure}

The functional form of $f_z$ and $f_M$ are completely arbitrary. The functions and values are chosen to match the four criteria mentioned above. We want to emphasise that a priori no physical argument was choosen to justify our approach except their result on the observables mentioned above. Recently \citep[][]{Oppenheimer:2008p6820,Oppenheimer:2010p6858,Henriques:2013p2980} provided physical pictures or reincorporation of gas and \citep[e.g,][]{Schaye:2010p5795} discussed extensively the impact of different physical processes on the evolution of the SFRD. The SFRD of Model C is plotted in Figure \ref{fig:SFRD_tuned} in red. We get about a factor of ten difference in the SFRD at $z=2$ and at $z=0$. The sSFR gets an additional boost at $z=2$ (red line in Figure \ref{fig:sSFR_tuned}) and the SMF at $z=0$ does match very well all the different galaxy populations in shape and amplitude (Figure \ref{fig:SMF_model_C}). The resulting SHMR plotted in Figure \ref{fig:SHMR_model_C} looks very different. The blue population is approaching the cosmic baryonic fraction very rapidly but gets quenched just before exceeding the limit (in stellar mass).

This extension can not be considered as a "best fit" model. The aim is just to indicate the power of this specific extension for future model buildings. Other predictions such as the gas-to-star ratio are only marginally affected by this extension.
We will not break the degeneracy between recycled and newly infallen gas components with this extension of our model. Metallicity and HI data \citep[see e.g, model of][]{Dave:2013p62} might give further insights into this processes.

\subsection{The coincidence of getting quenched when approaching the baryonic fraction} \label{sec:coincidence}

We notice from our analysis in Section \ref{sec:sh-relation} and \ref{sec:new_infall}, the SHMR is far below the cosmic baryonic fraction $f_b$ at low $M_s$ and is coming closer to $f_b$ when approaching $M^*$. By "coincidence", quenching occurs in our model just when the stellar baryonic fraction approaches the cosmic fraction $f_b$. In our model, the regulator is not allowed to get more baryons in than the baryonic fraction (see Equation \ref{eqn:infall})and so will automatically saturate. It will no longer follow the power law description of Section \ref{sec:sh-relation} and will flatten. In our model this saturation feature is completely independent of the quenching formalism with its crucial parameter $M^*$. 

However, apparently as a "coincidence", these two completely different features arise at the same point in the evolution history of a star-forming galaxy. It is ultimately this simultaneous appearance of these two features that led to the under-prediction of the red population around $M^*$. In our Model C, we see that to match the SMF we even have to steepen the SHMR of the blue population around $M^*$ such that the blue population must approach the cosmic baryonic limit even faster, without apparently noticing it, but suddenly then quench just before reaching the ultimate limit.

If one has one mechanism suppressing star formation in low mass galaxies and quenching at high masses, a peak is inevitable. But the peak in $M_s/M_h$ that is caused by quenching could have occurred at any mass, e.g. if it was driven by AGN feedback, morphological effects and so on. The fact that it appears to occur just when the overall efficiency of converting of baryons into stars is maximal is, in our view, noteworthy and probably tells us that it is not a coincidence.

\subsection{Abundance matching} \label{sec:abundance_matching}

We note from Figure \ref{fig:sh_scatter} (for Model A) and from Figure \ref{fig:SHMR_model_B} and \ref{fig:SHMR_model_C} (for Model B and C) that, at halo masses around $10^{12}M_{\odot}$, the mean value of the SHMR of the blue population is elevated by about 0.2 dex compared to the mean value of the red population. The 1-$\sigma$ dispersion in the blue population alone is about 0.2 dex, and the overall scatter in the combined red and blue populations at $10^{12}M_{\odot}$ is larger, 0.35 dex, and the distribution is not Gaussian in log $M_s/M_h$, i.e. log-normal in the ratio.  Simple abundance matching techniques usually do not take into account this possible variation.

\cite{Behroozi:2013p7464} noted that the range in star formation rates that is implicit in a star-forming and a passive population, is only a problem if it results in a distribution of stellar masses at fixed halo mass that cannot be reasonably modeled by a log-normal distribution (the main assumption in their work). In our particular model, we produce a clearly different distribution in stellar mass around the peak $M_s/M_h$. The SHMR of our Model C in Figure \ref{fig:SHMR_model_C} is substantially different to the one of \cite{Behroozi:2013p7464} but reproduces the SMF at the same accuracy.  In other words, the SHMR from our Model C is effectively a kind of abundance matching, as it is specifically tuned to match abundance properties of the galaxy population, but with a different assumption (motivated by our quenching laws) of how blue and red galaxies will populate the dark matter haloes.

\cite{Tinker:2013p7227} uses measurements of the stellar mass function, galaxy clustering, and galaxy-galaxy lensing within the COSMOS survey to constrain the SHMR of blue and red galaxies over the redshift range $z=[0.2,1]$. Their underlining assumption on the functional form of the blue and red galaxy SHMR is very different to our output. E.g. their blue population itselfs is described with a turn over in the SHMR.

\section{Conclusion}
\label{sec:conclusion}

We have presented a simple model of the evolving galaxy population that is based on importing pre-formulated baryonic prescriptions for the control of star-formation in galaxies into a dark matter halo merger tree. Specifically, the model is based on the gas-regulation model of star-forming galaxies from L13, and the empirical quenching formulae of P10 and P12.

The parameters for these baryonic prescriptions are taken directly from these earlier works and are not adjusted according to the output of the current model. A very limited number of additional a priori assumptions are however required to ensure the model can operate, but these do not greatly affect the outcome. The model allows us to make predictions about the numbers and properties of galaxies that are independent of the observation inputs used to determine the model prescriptions in the previous papers and which can therefore be used to test the model. The observational inputs to the previously tuned parameters were: The exponential cutoff scale $M^*$ of the main sequence galaxies at z=0, Z($M_s$,SFR) data at z=0 and the averaged enhanced fraction of red galaxies in groups and clusters. The only input from the dark matter picture in the previouse papers (namely L13) was the average halo growth rate of \cite{Neistein:2008p335}.

The output of this model is compared with independent observational data and also with other recent phenomenological models \cite{Behroozi:2013p7464}
for the evolving galaxy population that have been based on epoch-dependent abundance matching of haloes and galaxies. Output quantities examined include (a) the Main Sequence sSFR-mass relation; (b) the integrated star-formation rate density (SFRD); (c) the stellar mass functions of star-forming and quenched galaxies; (d) the $M_s vs. M_h$ relation and $SFR-M_h$ relations as well as the epoch dependence of these over the whole redshift interval $0 < z < 5$. The predicted gas content of galaxies is also presented.

The goal of this work has been to see how far we can get with this simple model and to explore how it may need to be adjusted so as to rectify any failings in reproducing the real Universe.

We have drawn the following conclusions out of this work:\\
\begin{enumerate}

	\item Reassuringly, the attractive features of the input baryonic prescriptions that were highlighted in the original papers, including the mass-dependence of the Main Sequence sSFR, the faint end slope of the galaxy mass function, the relative Schechter $M^*$ and $\alpha$ parameters of the blue and red (star-forming and quiescent) galaxy populations are certainly all preserved when transplanted into a realistic dark matter structure. The argument of L13 in relating the faint end slope $\alpha$ from the regulator scaling laws does not suffer from the limitations of a single mean sMIR. The mass-function of star-forming galaxies is also well reproduced and the general form of the $SFR-M_h$ and $M_s - M_h$ relations are very similar to those constructed by \cite{Behroozi:2013p7464} and arise from the competition between the increased efficiency of turning baryons into stars as the mass increases (due to lower mass loss in winds) and the quenching of star-formation in galaxies. The overall forms of the sSFR(z) and SFRD(z) are also qualitatively produced by the model. These are major and rather striking successes from an simple model that are very largely independent of the original observational inputs that were used previously to define our baryonic prescriptions.
	
	\item As with other models in the literature, our simplest model has quantitative difficulty in reproducing the steep increase back to $z \sim 2$ in both the sSFR(z) and SFRD(z). This cannot be solved by simple adjustments to the adopted star-formation efficiencies. We also find that the peak in the $M_s - M_h$ relation is a little softer than in the \cite{Behroozi:2013p7464} representation and, surprisingly, that the ratio of quenched to star-forming galaxies around $M*$ is lower than observed (and than can be predicted from the original P10 formalism). We show that the latter two issues are closely related and are due to a saturation in the efficiency with which haloes form stars that is inherent in the adopted regulator model, especially as the cosmic baryon limit is approached. 
		
	\item All four of these quantitative deficiencies can be simultaneously solved by adjusting the specific infall rate of material onto galaxies by allowing them to re-ingest material previously expelled by winds provided that this occurs in a redshift- and mass-dependent way, being most effective at masses around $M^*$ and at redshifts $z \sim 2$. 
		
	\item Our model allow us to predict the $M_s - M_h$ relation for star-forming and quiescent galaxies separately. Red galaxies always have a higher $M_h$ at given $M_s$ because of the continued growth of haloes after star-formation ceases, and there is a 1-$\sigma$ scatter in stellar mass of 0.36 dex for haloes of mass $M_h=10^{12}M_{\odot}$ with two clearly distinguishable populations. There is significantly less scatter in the blue population than in the red one. The SHMR around $M^*$, where mass quenching happens, has to be steeper than predicted from our original model to match the blue and red galaxy abundances at the same time. Such a qualitative behavior brings a simple regulator model to its limits as one expects the SHMR to flatten when approaching the baryonic limit. 

	\item While others have emphasized the ``inefficiency'' of star-formation in haloes, we stress instead the efficiency of $M^*$ galaxies in forming stars. Further, we note the "coincidence" that quenching happens in our model just at the time when the regulators are rapidly approaching the maximum possible efficiency in covering baryons into stars, even thought these two description are completely independent of each other in the model.
	
\end{enumerate}

Our analysis emphasizes the continued importance of pinning down as reliably as possible the bulk characteristics of the evolving galaxy population over a wide span of cosmic time. One crucial factor in our model is the gas infall onto galaxies, and it will be of great importance to trace the gas in the universe in a more observationally comprehensive way.

\section*{Acknowledgment}
We thank Neven Caplar and Sebastian Seehars for helping us doing the Schechter fits of our model samples.
SJL thanks his co-authors on the P10, P12 and L13 papers for their indirect but nevertheless substantial contributions to the present work. SB thanks Sandro Tacchella for usefull discussion for the comparision with data samples.
This work has been supported by the Swiss National Science Foundation (grant number 200021-143906/1 and 200020-140683).

\bibliography{papers_bibtex}{}

\appendix

\section{Model sensitivity on additional parameters and initial conditions} \label{ap:sensitivity}

\subsection{Model dependence on $f_{\text{merge}}$} \label{ap:merging}

\begin{figure}
  \centering
  \includegraphics[angle=0, width=80mm]{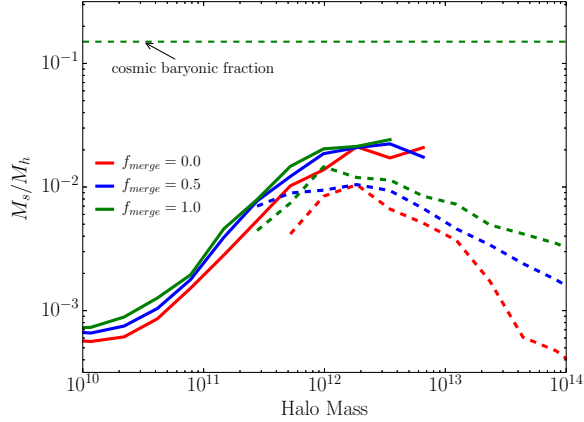}
  \caption{The SHMR at $z=0$ is plotted for three different $f_{\text{merge}}$. Filled lines indicate the mean in the blue population and dotted lines the mean of the red populations. Increasing values for $f_{\text{merge}}$ do only significantly affect the high halo mass end.}
  \label{fig:f_merge_compare}
\end{figure}

The parameter $f_{\text{merge}}$ in the model describes the fraction of the stellar and gas mass of a satellite that enters the central galaxy when the satellite is disrupted.   In the main text, this is set to $f_{\text{merge}}=0.5$.  To understand the sensitivity to this parameter we present here three models that each have the same parameters as in Table \ref{tab:parameters} except that $f_{\text{merge}}$ is set to $f_{\text{merge}}=0, 0.5, 1$.  There is a small change in the blue population.  Even with $f_{\text{merge}}=1$, a blue central galaxy almost never has more than 20\% of it's mass growth through merging.   For the red galaxies at $m_s >> M*$, however, which are generally located in massive haloes (e.g. above $10^{13}M_{\odot}$), merging is the primary channel for mass growth and there is therefore a significant effect of $f_{merge}$ on the the mass function and the SHMR for these most massive galaxies.  The effect on the SHMR  is shown in Figure \ref{fig:f_merge_compare}. If we want to predict this quantity of the stellar mass function at these high masses, then we would have to constrain $f_{\text{merge}}$ (or vice versa). This regime of galaxy mass is not however a central consideration of this paper.

\subsection{Model dependence on $\lambda_{\text{max}}$} \label{ap:outflow}

In the model, $\lambda_{\text{max}}$ gives the maximum mass-loading of the wind, which is required to limit the extrapolation of the $\lambda$ which varies inversely with mass at higher masses. In Figure \ref{fig:floor_compare}, three models with different $\lambda_{\text{max}}$ are plotted. As would be expected, there is a significant dependence for the lowest mass galaxies, corresponding to haloes below $10^{11}M_{\odot}$, where the SHMR scales linearly with $\lambda_{\text{max}}^{-1}$ at the very low stellar mass end. We choose $\lambda_{\text{max}}$ in such a way that the regulator above $10^{9}M_{\odot}$ in stellar mass is not affected by the floor value. This gives us a prior of $\lambda_{\text{max}}\geq20$. Values between 20-200 only affect the intermediate range marginally. We choose $\lambda_{\text{max}}=50$ and note that our model is not tuned to predict the SHMR below $10^{11}M_{\odot}$ in halo mass or $10^{8}M_{\odot}$ in stellar mass correspondingly.

\subsection{Model dependence on $M_{\text{thresh}}$} \label{ap:m_thresh}

\begin{figure}[htbp]
  \begin{minipage}[b]{0.5\linewidth}
    \centering
	\includegraphics[angle=0, width=80mm]{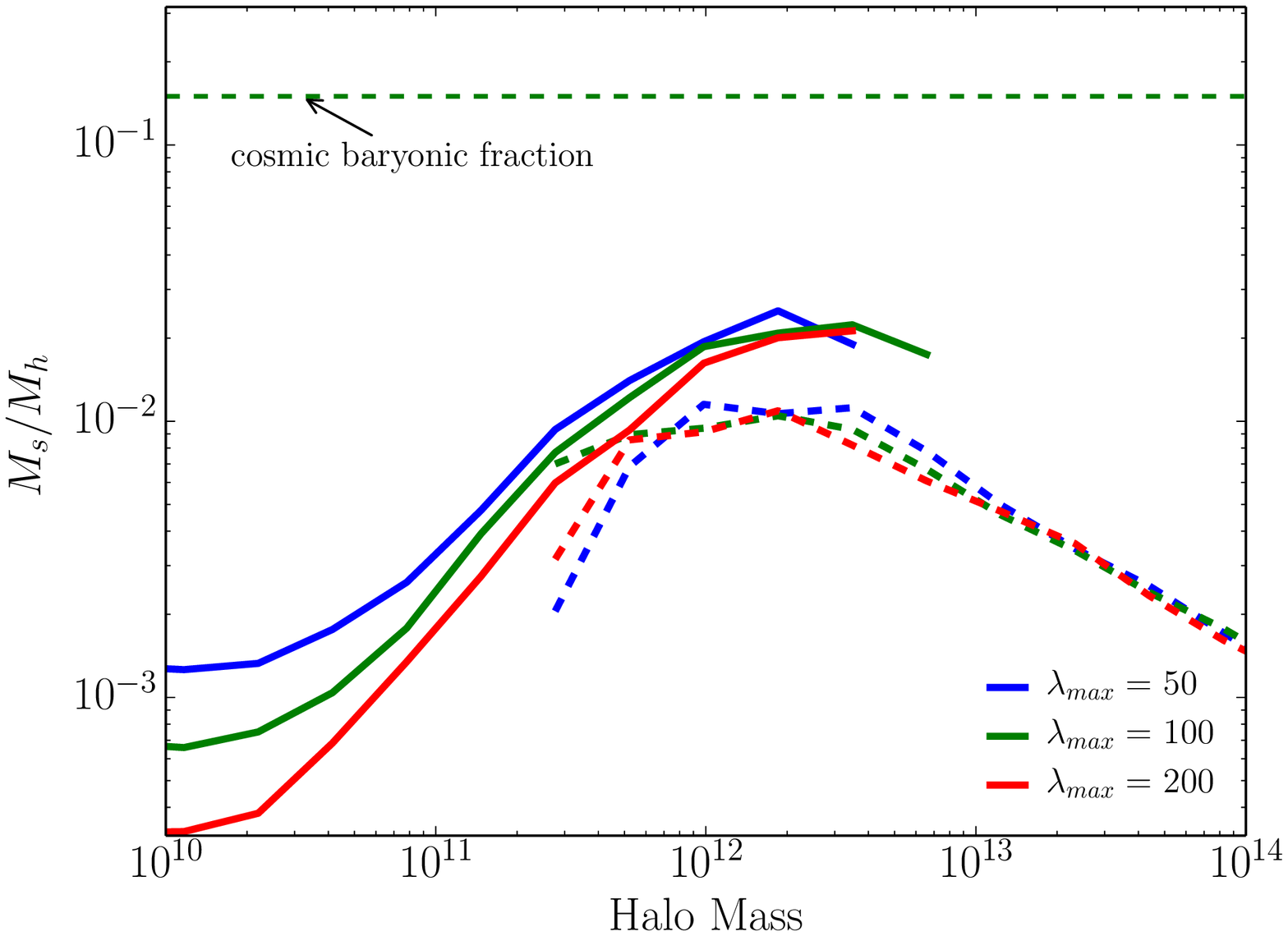}
	\caption{The SHMR at $z=0$ is plotted for three different $\lambda_{\text{max}}$. In color are the mean values of the blue population. The mean values of the red fraction are plotted in red for all four models. We detect no significant deviation in the red fraction above the turn-over. The low mass population are significantly affected by $\lambda_{\text{max}}$.}
	\label{fig:floor_compare}
  \end{minipage}
  \hspace{0.5cm}
  \begin{minipage}[b]{0.5\linewidth}
    \centering
  	\includegraphics[angle=0, width=80mm]{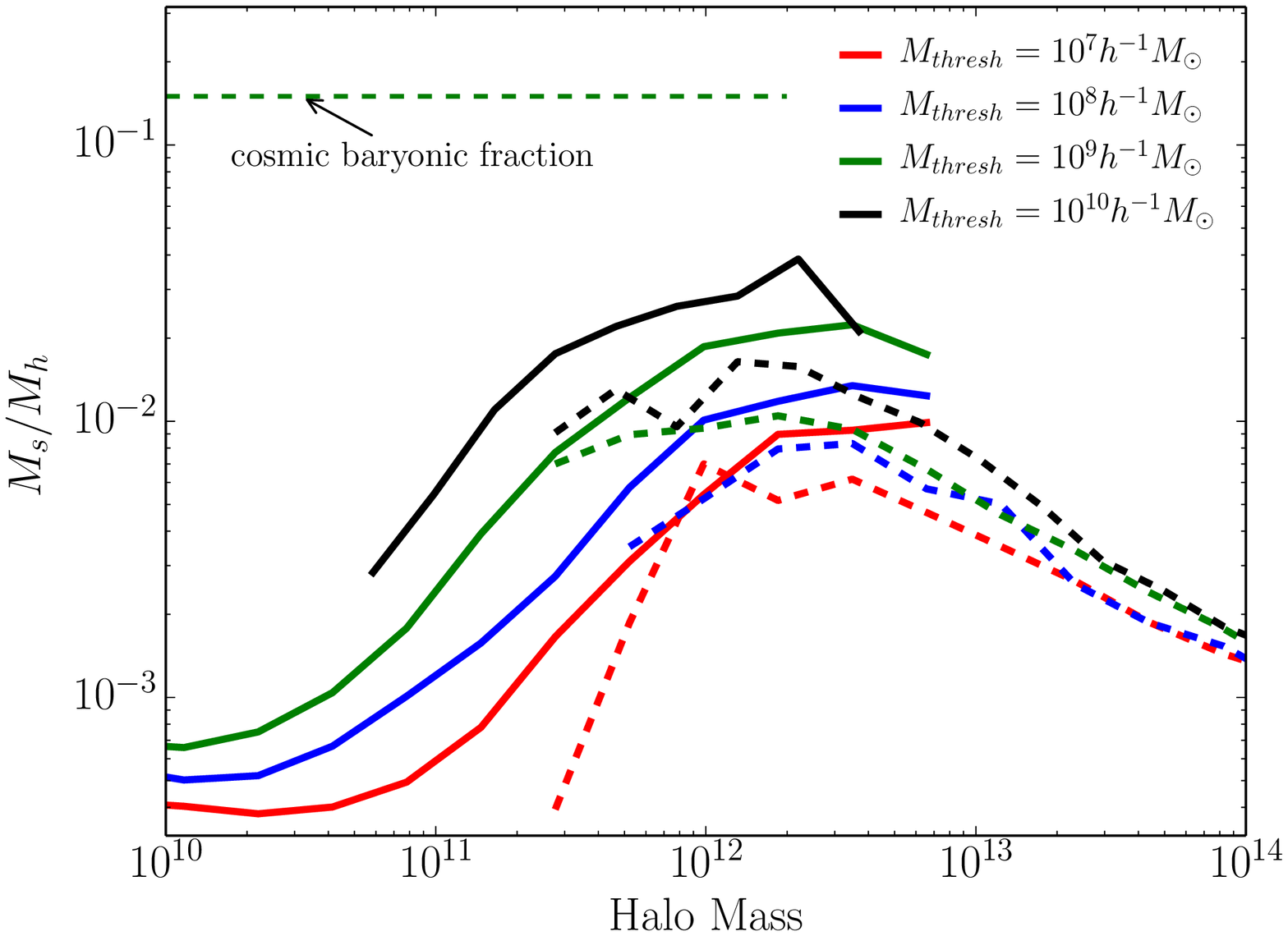}
  	\caption{The SHMR at $z=0$ is plotted for four different $M_{\text{thresh}}$. In color are the mean values of the blue population. The mean values of the red fraction are plotted in red for all four models. When changing $M_{\text{thresh}}$ over four orders of magnitude, the SHMR changes by less than one order of magnitude.}
  	\label{fig:m_thresh_compare}
  \end{minipage}
\end{figure}

The parameter $M_{\text{thresh}}$ controls the threshold above which a (sub-) halo contains a regulator system, which in turn affects the way in which baryons are brought into the larger haloes.  The dependence of our model on $M_{\text{thresh}}$ parameter is a little more complicated. First, when lowering $M_{\text{thresh}}$ we increase the merging component $\dot{M}_{\text{merger}}$ and lower the smoothed accretion component $\dot{M}_{\text{smoothed}}$. Second, the very high host-to-satellite ratio makes those low mass substructures survive very long (often longer than the age of the Universe). These results that $M_{\text{central}}$ defined by Equation (\ref{eqn:central}) is lowered compared to $M_h$. From Equation (\ref{eqn:infall}), a lowered $M_{\text{central}}$ leads to a reduction of the infall rate in Model A. In Figure \ref{fig:m_thresh_compare} four models with different $M_{\text{thresh}}$ are plotted. We see a scaling difference in the SHMR. When changing $M_{\text{thresh}}$ by four orders of magnitude, we change the SHMR by less than one order of magnitude.  L13 introduced in their paper a parameter $f_{\text{gal}}$ to account for the fact that if they let all the accreted baryons in their regulator they would end up with to high an SHMR. In our model, we naturally do not let all the baryons fall in the central because of the sub-structure. The parameter $f_{\text{gal}}$ is therefore effectively absorbed into the (more physical) parameter $M_{\text{thresh}}$. We get an equivalent of $f_{\text{gal}}=0.5$ with $M_{\text{thresh}}=10^9h^{-1}M_{\odot}$. This mass is consistent with photo-ionisation heating operating at low masses and suppress cooling and star formation below a certain halo mass $M_{\gamma}$. This halo mass scale increases from $M_{\gamma}\sim 10^8M_{\odot}$ during reionisation to $M_{\gamma}\sim \text{few}\cdot 10^9M_{\odot}$ \citep[][]{Gnedin:2000p1139,Okamoto:2008p1151}. For a more realistic model aiming to make predictions back to the epoch of reionisation, one has to account for a change in the mass threshold.

\subsection{Initial conditions} \label{ap:init_cond}

When we stop expanding our merger tree (in a backward process) at either redshift $z=15$ or at halo masses of $10^9h^{-1}M_{\odot} < M_h < 2\cdot10^9h^{-1}M_{\odot}$ we have to initialize the baryonic component of the halo. To start the forward process of the regulator system, we have to put in some initial values for $M_s$ and $M_{\text{gas}}$. In principle we should start with $M_{s,\text{init}}=0$. But with this initial condition the star-formation efficiency is zero and so the differential equation we want to solve has the solution $M_s(t)=0$ for all times. Whether we start with $M_{s\text{ init}}=1M_{\odot}$ or $M_{s\text{ init}}=10^3M_{\odot}$ does not really matter when predicting the quantities in $M_s=10^8M_{\odot}$ galaxies. The time to form these first $10^3M_{\odot}$ is rather short when considering a gas reservoir of order $M_{\text{gas}} \approx 10^8M_{\odot}$. For the initial condition of the gas content in the regulator $M_{\text{gas,init}}$ we have the freedom of $0 < M_{\text{gas,init}} < f_b M_{h,\text{init}}$. This has not more than a 1\% effect on the total amount of gas that comes into a halo of mass $M_h=10^{11}M_{\odot}$. We conclude that for merger trees which reach $M_{\text{thresh}}$, our freedom in the initial condition do not affect our predictions by more than 1\%.

The situation for starting the forward process at $z = 15$ with a halo more massive than $M_{\text{thresh}}$ is slightly different. This case only happens for haloes of present-day mass $M_h > 10^{14}M_{\odot}$. We cannot a priori predict the stellar or gas content of a halo of $M_h=10^{12}M_{\odot}$ at $z=15$.  However, we study the output of the model only at $z < 8$ , by which point these haloes have grown in mass by an order of magnitude. Whatever initial conditions we put in, it affects predictions at $z = 8$ by only about 10\%, and even less at later epochs.

\section{Tables and fits} \label{ap:tables}

In this section we provide tables with functional fits for the mass functions from the simple models presented in the paper and for the SHMR at $z=0$.  As discussed in the main text, the goal of this work has not been to perfectly match observational data, but rather to explore the consequences of particularly simple representations of galaxy evolution.

\subsection{Stellar mass function}

We have fitted Schechter functions to the blue central, red central, blue satellite and red satellite galaxy population. Fits are made assuming a fixed 10\% error in log space for each binning point of the sample of Section \ref{sec:results} plotted in Figure \ref{fig:smf_evolution}. We define our parameters according to
\begin{equation} \label{eqn:schechter}
	\phi(m)\text{d}m=\phi_*\left( \frac{m}{M^*}\right)^{\alpha_s}e^{-m/M^*}\frac{\text{d}m}{M^*}
\end{equation}
with $\phi$ being the number density in units of Mpc$^{-3}/\text{d}m$ and $M^*$ in units of $M_{\odot}$. Fits are made over all stellar masses above $10^8M_{\odot}$. The fitted functions were integrated through the bins to compare with the number of galaxies in the model.
The fitted Schechter parameters and their errors are given in Table \ref{tab:smf_blue_central} for blue centrals, Table \ref{tab:smf_blue_satellites} for the blue satellites, Table \ref{tab:smf_red_centrals} for the red centrals and Table \ref{tab:smf_red_satellites} for the red satellite population for Model A. Fits for Model B are provided in Table \ref{tab:smf_blue_central_B}-\ref{tab:smf_red_satellites_B} and for Model C in Table \ref{tab:smf_blue_centrals_C}-\ref{tab:smf_red_satellites_C}. 

It should be noted that the fitted $M^*$ from the model output(s) are not equivalent to the model parameter $M^*$ that is used in the paper.  Merging after quenching will result in a higher $M^*$ fit.  Also, at high redshifts, the exponential cut-off of the Press-Schechter-like dark matter halo mass function results in a lower value of the Schechter function $M^*$ during Phase 1 in the parlance of \cite{Lilly:2013p7458}.

In fact, our fits at redshifts above $z \sim 3$ are influenced by the Press-Schechter shape of the stellar mass function and the fits might be not as good as at lower redshifts. According to P12, the red satellite population has the form of a double-Schechter function. When fitting a double-Schechter function to our red satellite population, the mass-quenched part of the Schechter function cannot be well constrained and a single Schechter function provides a reasonable fit to our sample of red satellites for most redshifts. Note however that this does not invalidate the explanation of P12 and the implied difference in $\alpha$.  We provide in Table \ref{tab:smf_red_satellites} and \ref{tab:smf_red_satellites_B} these single Schechter fits.

\begin{deluxetable*}{rrrrrrr}
\tabletypesize{\scriptsize}
\tablecaption{This table lists Schechter fits of the sample plotted in Figure \ref{fig:smf_evolution} for blue centrals of Model A. Parameterization is according to Equation (\ref{eqn:schechter}).}
\tablewidth{0pt}
\tablehead{z	 & log$_{10}$($\phi_{*}$) [Mpc$^{-3}$]	& log$_{10}$($\phi_{*}$) [Mpc$^{-3}$] Error &	log$_{10}$$(M^*/M_{\odot})$	&		log$_{10}(M^*/M_{\odot})$ Error	&	$\alpha_s$	&	$\alpha_s$ Error}

0.0 & -2.98 & 0.04 & 10.65 & 0.021 & -1.48 & 0.018 \\
0.5 & -3.06 & 0.03 & 10.64 & 0.018 & -1.51 & 0.017 \\
1.0 & -3.09 & 0.04 & 10.57 & 0.019 & -1.51 & 0.02 \\
1.5 & -3.29 & 0.04 & 10.59 & 0.02 & -1.57 & 0.019 \\
2.0 & -3.51 & 0.04 & 10.59 & 0.018 & -1.64 & 0.017 \\
2.5 & -3.63 & 0.03 & 10.51 & 0.016 & -1.67 & 0.016 \\
3.0 & -3.8 & 0.03 & 10.44 & 0.012 & -1.72 & 0.017 \\
3.5 & -4.15 & 0.04 & 10.44 & 0.013 & -1.83 & 0.019 \\
4.0 & -4.25 & 0.03 & 10.32 & 0.01 & -1.84 & 0.02 \\
4.5 & -4.8 & 0.04 & 10.34 & 0.011 & -2.05 & 0.02 \\
5.0 & -5.19 & 0.05 & 10.33 & 0.014 & -2.16 & 0.022 \\
5.5 & -5.7 & 0.04 & 10.32 & 0.014 & -2.31 & 0.02 \\
6.0 & -5.64 & 0.03 & 10.06 & 0.008 & -2.3 & 0.021 \\

\label{tab:smf_blue_central}
\end{deluxetable*}

\begin{deluxetable*}{rrrrrrr}
\tabletypesize{\scriptsize}
\tablecaption{This table lists Schechter fits of the sample plotted in figure \ref{fig:smf_evolution} for blue satellites of Model A. Parameterization is according to equation (\ref{eqn:schechter}).}
\tablewidth{0pt}
\tablehead{z	 & log$_{10}$($\phi_{*}$) [Mpc$^{-3}$]	& log$_{10}$($\phi_{*}$) [Mpc$^{-3}$] Error &	log$_{10}$$(M^*/M_{\odot})$	&		log$_{10}(M^*/M_{\odot})$ Error	&	$\alpha_s$	&	$\alpha_s$ Error}

0.0 & -3.96 & 0.04 & 10.55 & 0.019 & -1.53 & 0.019 \\
0.5 & -4.2 & 0.04 & 10.63 & 0.021 & -1.6 & 0.018 \\
1.0 & -4.34 & 0.04 & 10.57 & 0.019 & -1.65 & 0.017 \\
1.5 & -4.63 & 0.25 & 10.59 & 0.166 & -1.72 & 0.037 \\
2.0 & -4.67 & 0.03 & 10.44 & 0.012 & -1.73 & 0.018 \\
2.5 & -4.96 & 0.05 & 10.42 & 0.02 & -1.79 & 0.022 \\
3.0 & -5.27 & 0.04 & 10.38 & 0.015 & -1.86 & 0.02 \\
3.5 & -5.91 & 0.06 & 10.48 & 0.022 & -2.04 & 0.022 \\
4.0 & -6.05 & 0.04 & 10.25 & 0.014 & -2.12 & 0.022 \\
4.5 & -5.78 & 0.05 & 9.889 & 0.016 & -2.01 & 0.028 \\
5.0 & -6.67 & 5.76 & 10.16 & 4.808 & -2.19 & 0.029 \\
5.5 & -6.01 & 0.38 & 9.432 & 0.153 & -2.02 & 0.095 \\
6.0 & -6.68 & 0.29 & 9.455 & 0.133 & -2.31 & 0.072 \\

\label{tab:smf_blue_satellites}
\end{deluxetable*}

\begin{deluxetable*}{rrrrrrr}
\tabletypesize{\scriptsize}
\tablecaption{This table lists Schechter fits of the sample plotted in Figure \ref{fig:smf_evolution} for red centrals of Model A. Parameterization is according to Equation (\ref{eqn:schechter}).}
\tablewidth{0pt}
\tablehead{z	 & log$_{10}$($\phi_{*}$) [Mpc$^{-3}$]	& log$_{10}$($\phi_{*}$) [Mpc$^{-3}$] Error &	log$_{10}$$(M^*/M_{\odot})$	&		log$_{10}(M^*/M_{\odot})$ Error	&	$\alpha_s$	&	$\alpha_s$ Error}

0.0 & -2.77 & 0.02 & 10.58 & 0.017 & -0.33 & 0.017 \\
0.5 & -2.92 & 0.02 & 10.56 & 0.018 & -0.4 & 0.019 \\
1.0 & -3.13 & 0.02 & 10.57 & 0.016 & -0.54 & 0.017 \\
1.5 & -3.33 & 0.02 & 10.54 & 0.018 & -0.6 & 0.019 \\
2.0 & -3.52 & 0.02 & 10.48 & 0.014 & -0.58 & 0.018 \\
2.5 & -3.79 & 0.02 & 10.44 & 0.014 & -0.67 & 0.018 \\
3.0 & -4.09 & 0.03 & 10.41 & 0.014 & -0.72 & 0.02 \\
3.5 & -4.5 & 0.02 & 10.41 & 0.012 & -0.88 & 0.019 \\
4.0 & -4.88 & 0.02 & 10.33 & 0.011 & -1.01 & 0.018 \\
4.5 & -5.33 & 0.03 & 10.29 & 0.013 & -1.07 & 0.02 \\
5.0 & -5.81 & 0.03 & 10.26 & 0.012 & -1.21 & 0.02 \\
5.5 & -6.24 & 0.44 & 10.12 & 0.216 & -1.3 & 0.092 \\
6.0 & -7.11 & 0.04 & 10.26 & 0.014 & -1.66 & 0.022 \\

\label{tab:smf_red_centrals}
\end{deluxetable*}

\begin{deluxetable*}{rrrrrrr}
\tabletypesize{\scriptsize}
\tablecaption{This table lists Schechter fits of the sample plotted in figure \ref{fig:smf_evolution} for red satellites of Model A. Parameterization is according to equation (\ref{eqn:schechter}).}
\tablewidth{0pt}
\tablehead{z	 & log$_{10}$($\phi_{*}$) [Mpc$^{-3}$]	& log$_{10}$($\phi_{*}$) [Mpc$^{-3}$] Error &	log$_{10}$$(M^*/M_{\odot})$	&		log$_{10}(M^*/M_{\odot})$ Error	&	$\alpha_s$	&	$\alpha_s$ Error}

0.0 & -4.24 & 0.13 & 10.64 & 0.038 & -1.54 & 0.067  \\
0.5 & -4.17 & 0.05 & 10.76 & 0.064 & -1.46 & 0.029  \\
1.0 & -4.38 & 0.05 & 10.74 & 0.071 & -1.54 & 0.03  \\
1.5 & -4.64 & 0.04 & 10.75 & 0.039 & -1.6 & 0.02  \\
2.0 & -4.85 & 0.04 & 10.64 & 0.040 & -1.66 & 0.022  \\
2.5 & -5.19 & 0.07 & 10.59 & 0.035 & -1.75 & 0.041  \\
3.0 & -5.95 & 0.80 & 10.62 & 0.027 & -1.98 & 0.229  \\
3.5 & -6.00 & 0.72 & 10.40 & 0.049 & -2.0 & 0.058  \\
4.0 & -6.60 & 0.89 & 10.38 & 0.046 & -2.17 & 0.106  \\
4.5 & -6.55 & 0.51 & 10.16 & 0.045 & -2.17 & 0.057  \\
5.0 & -8.57 & 5.27 & 10.32 & 0.109 & -2.65 & 0.147  \\
5.5 & -10.06 & 4.34 & 10.8 & 0.280 & -2.64 & 0.187  \\
6.0 & -11.16 & 1.88 & 10.9 & 0.277 & -2.95 & 0.238  \\

\label{tab:smf_red_satellites}
\end{deluxetable*}

\begin{deluxetable*}{rrrrrrr}
\tabletypesize{\scriptsize}
\tablecaption{This table lists Schechter fits of the sample plotted of Model B for blue centrals. Parameterization is according to Equation (\ref{eqn:schechter}).}
\tablewidth{0pt}
\tablehead{z	 & log$_{10}$($\phi_{*}$) [Mpc$^{-3}$]	& log$_{10}$($\phi_{*}$) [Mpc$^{-3}$] Error &	log$_{10}$$(M^*/M_{\odot})$	&		log$_{10}(M^*/M_{\odot})$ Error	&	$\alpha_s$	&	$\alpha_s$ Error}
0.0 & -2.91 & 0.04 & 10.68 & 0.023 & -1.47 & 0.018 \\
0.5 & -2.94 & 0.04 & 10.64 & 0.022 & -1.47 & 0.019 \\
1.0 & -3.03 & 0.04 & 10.61 & 0.019 & -1.50 & 0.019 \\
1.5 & -3.18 & 0.04 & 10.63 & 0.020 & -1.54 & 0.017 \\
2.0 & -3.28 & 0.04 & 10.59 & 0.018 & -1.55 & 0.017 \\
2.5 & -3.47 & 0.04 & 10.58 & 0.019 & -1.61 & 0.018 \\
3.0 & -3.66 & 0.04 & 10.55 & 0.017 & -1.65 & 0.018 \\
3.5 & -3.96 & 0.04 & 10.56 & 0.017 & -1.74 & 0.018 \\
4.0 & -4.15 & 0.04 & 10.48 & 0.014 & -1.79 & 0.017 \\
4.5 & -4.32 & 0.04 & 10.39 & 0.013 & -1.83 & 0.019 \\
5.0 & -4.75 & 0.03 & 10.38 & 0.011 & -1.97 & 0.016 \\
5.5 & -5.16 & 0.04 & 10.36 & 0.016 & -2.08 & 0.020 \\
6.0 & -4.83 & 0.03 & 9.957 & 0.006 & -1.97 & 0.021 \\

\label{tab:smf_blue_central_B}
\end{deluxetable*}

\begin{deluxetable*}{rrrrrrr}
\tabletypesize{\scriptsize}
\tablecaption{This table lists Schechter fits of the sample of Model B for blue satellites. Parameterization is according to equation (\ref{eqn:schechter}).}
\tablewidth{0pt}
\tablehead{z	 & log$_{10}$($\phi_{*}$) [Mpc$^{-3}$]	& log$_{10}$($\phi_{*}$) [Mpc$^{-3}$] Error &	log$_{10}$$(M^*/M_{\odot})$	&		log$_{10}(M^*/M_{\odot})$ Error	&	$\alpha_s$	&	$\alpha_s$ Error}

0.0 & -3.98 & 0.04 & 10.61 & 0.022 & -1.45 & 0.020 \\
0.5 & -4.07 & 0.04 & 10.61 & 0.021 & -1.49 & 0.019 \\
1.0 & -4.09 & 0.03 & 10.52 & 0.015 & -1.5 & 0.017 \\
1.5 & -4.35 & 0.04 & 10.56 & 0.019 & -1.55 & 0.019 \\
2.0 & -4.50 & 0.04 & 10.5 & 0.016 & -1.59 & 0.019 \\
2.5 & -4.64 & 0.03 & 10.4 & 0.011 & -1.6 & 0.018 \\
3.0 & -5.22 & 0.04 & 10.47 & 0.014 & -1.82 & 0.018 \\
3.5 & -5.28 & 0.04 & 10.31 & 0.014 & -1.78 & 0.02 \\
4.0 & -5.71 & 0.04 & 10.25 & 0.013 & -1.92 & 0.022 \\
4.5 & -6.04 & 0.41 & 10.16 & 0.177 & -2.0 & 0.064 \\
5.0 & -7.07 & 0.1 & 10.35 & 0.047 & -2.33 & 0.029 \\
5.5 & -6.45 & 0.2 & 9.73 & 0.093 & -2.24 & 0.048 \\
6.0 & -6.65 & 0.07 & 9.687 & 0.026 & -2.02 & 0.037 \\

\label{tab:smf_blue_satellites_B}
\end{deluxetable*}

\begin{deluxetable*}{rrrrrrr}
\tabletypesize{\scriptsize}
\tablecaption{This table lists Schechter fits of the sample of Model B for red centrals. Parameterization is according to Equation (\ref{eqn:schechter}).}
\tablewidth{0pt}
\tablehead{z	 & log$_{10}$($\phi_{*}$) [Mpc$^{-3}$]	& log$_{10}$($\phi_{*}$) [Mpc$^{-3}$] Error &	log$_{10}$$(M^*/M_{\odot})$	&		log$_{10}(M^*/M_{\odot})$ Error	&	$\alpha_s$	&	$\alpha_s$ Error}

0.0 & -2.71 & 0.02 & 10.66 & 0.02 & -0.48 & 0.018 \\
0.5 & -2.86 & 0.03 & 10.67 & 0.022 & -0.58 & 0.019 \\
1.0 & -2.97 & 0.02 & 10.62 & 0.019 & -0.58 & 0.018 \\
1.5 & -3.16 & 0.02 & 10.61 & 0.019 & -0.66 & 0.018 \\
2.0 & -3.36 & 0.03 & 10.59 & 0.020 & -0.71 & 0.019 \\
2.5 & -3.56 & 0.02 & 10.53 & 0.015 & -0.71 & 0.019 \\
3.0 & -3.89 & 0.03 & 10.54 & 0.017 & -0.81 & 0.019 \\
3.5 & -4.19 & 0.03 & 10.48 & 0.014 & -0.9 & 0.019 \\
4.0 & -4.54 & 0.03 & 10.44 & 0.014 & -0.98 & 0.019 \\
4.5 & -4.93 & 0.03 & 10.41 & 0.013 & -1.07 & 0.02 \\
5.0 & -5.19 & 0.03 & 10.26 & 0.011 & -1.07 & 0.019 \\
5.5 & -5.43 & 0.03 & 10.05 & 0.008 & -1.06 & 0.024 \\
6.0 & -5.74 & 0.03 & 9.83 & 0.013 & -1.08 & 0.027 \\
\label{tab:smf_red_centrals_B}
\end{deluxetable*}

\begin{deluxetable*}{rrrrrrr}
\tabletypesize{\scriptsize}
\tablecaption{This table lists Schechter fits of the sample of Model B for red satellites. Parameterization is according to equation (\ref{eqn:schechter}).}
\tablewidth{0pt}
\tablehead{z	 & log$_{10}$($\phi_{*}$) [Mpc$^{-3}$]	& log$_{10}$($\phi_{*}$) [Mpc$^{-3}$] Error &	log$_{10}$$(M^*/M_{\odot})$	&		log$_{10}(M^*/M_{\odot})$ Error	&	$\alpha_s$	&	$\alpha_s$ Error}

0.0 & -3.99 & 0.08 & 10.70 & 0.054 & -1.44 & 0.046  \\
0.5 & -4.04 & 0.07 & 10.69 & 0.053 & -1.45 & 0.041  \\
1.0 & -4.20 & 0.07 & 10.63 & 0.041 & -1.5 & 0.039  \\
1.5 & -4.37 & 0.05 & 10.69 & 0.067 & -1.52 & 0.031  \\
2.0 & -4.68 & 0.04 & 10.75 & 0.048 & -1.59 & 0.022  \\
2.5 & -4.91 & 0.04 & 10.64 & 0.039 & -1.65 & 0.021  \\
3.0 & -5.29 & 0.45 & 10.61 & 0.030 & -1.74 & 0.107  \\
3.5 & -5.67 & 0.30 & 10.55 & 0.034 & -1.84 & 0.058  \\
4.0 & -6.42 & 1.28 & 10.52 & 0.045 & -2.06 & 0.099  \\
4.5 & -6.80 & 1.00 & 10.39 & 0.062 & -2.19 & 0.177  \\
5.0 & -6.57 & 1.24 & 9.945 & 0.042 & -2.11 & 0.155  \\
5.5 & -6.76 & 1.33 & 9.81 & 0.111 & -2.19 & 0.163  \\
6.0 & -5.94 & 1.38 & 9.135 & 0.63 & -1.53 & 0.452  \\

\label{tab:smf_red_satellites_B}
\end{deluxetable*}

\begin{deluxetable*}{rrrrrrr}
\tabletypesize{\scriptsize}
\tablecaption{This table lists Schechter fits of the sample of Model C for blue centrals. Parameterization is according to equation (\ref{eqn:schechter}).}
\tablewidth{0pt}
\tablehead{z	 & log$_{10}$($\phi_{*}$) [Mpc$^{-3}$]	& log$_{10}$($\phi_{*}$) [Mpc$^{-3}$] Error &	log$_{10}$$(M^*/M_{\odot})$	&		log$_{10}(M^*/M_{\odot})$ Error	&	$\alpha_s$	&	$\alpha_s$ Error}

0.0	&-3.21	&0.04	&10.71	&0.025	&-1.49	&0.019\\
0.5	&-3.1	&0.04	&10.59	&0.019	&-1.45	&0.019\\
1.0	&-3.26	&0.04	&10.66	&0.024	&-1.47	&0.02\\
1.5	&-3.48	&0.04	&10.74	&0.026	&-1.51	&0.019\\
2.0	&-3.61	&0.04	&10.7	&0.022	&-1.51	&0.018\\
2.5	&-3.95	&0.04	&10.8	&0.029	&-1.55	&0.016\\
3.0	&-4.23	&0.05	&10.83	&0.03	&-1.58	&0.018\\
3.5	&-4.45	&0.05	&10.76	&0.029	&-1.6	&0.019\\
4.0	&-4.62	&0.04	&10.68	&0.023	&-1.59	&0.019\\
4.5	&-5.01	&0.05	&10.72	&0.027	&-1.65	&0.019\\
5.0	&-5.12	&0.04	&10.55	&0.018	&-1.61	&0.021\\
5.5	&-5.93	&0.07	&10.76	&0.041	&-1.8	&0.019\\
6.0	&-6.48	&0.06	&10.69	&0.032	&-1.94	&0.021\\
\label{tab:smf_blue_centrals_C}
\end{deluxetable*}

\begin{deluxetable*}{rrrrrrr}
\tabletypesize{\scriptsize}
\tablecaption{This table lists Schechter fits of the sample of Model c for red centrals. Parameterization is according to Equation (\ref{eqn:schechter}).}
\tablewidth{0pt}
\tablehead{z	 & log$_{10}$($\phi_{*}$) [Mpc$^{-3}$]	& log$_{10}$($\phi_{*}$) [Mpc$^{-3}$] Error &	log$_{10}$$(M^*/M_{\odot})$	&		log$_{10}(M^*/M_{\odot})$ Error	&	$\alpha_s$	&	$\alpha_s$ Error}
0.0&	-2.47	&0.02	&10.61	&0.02	&-0.12	&0.018\\
0.5&	-2.55	&0.02	&10.65	&0.019	&-0.2	&0.018\\
1.0&	-2.61	&0.02	&10.62	&0.018	&-0.13	&0.018\\
1.5&	-2.86	&0.02	&10.73	&0.023	&-0.34	&0.017\\
2.0&	-3.07	&0.02	&10.73	&0.026	&-0.37	&0.02\\
2.5&	-3.35	&0.02	&10.75	&0.026	&-0.43	&0.019\\
3.0&	-3.64	&0.02	&10.75	&0.026	&-0.46	&0.018\\
3.5&	-3.98	&0.02	&10.77	&0.025	&-0.53	&0.018\\
4.0&	-4.32	&0.02	&10.74	&0.026	&-0.57	&0.02\\
4.5&	-4.81	&0.03	&10.84	&0.032	&-0.73	&0.018\\
5.0&	-5.32	&0.03	&10.82	&0.032	&-0.85	&0.019\\
5.5&	-5.96	&0.04	&10.87	&0.036	&-1.03	&0.019\\
6.0&	-6.61	&0.06	&10.93	&0.763	&-1.12	&0.02\\

\label{tab:smf_red_centrals_C}
\end{deluxetable*}

\begin{deluxetable*}{rrrrrrr}
\tabletypesize{\scriptsize}
\tablecaption{This table lists Schechter fits of the sample of Model C for blue satellites. Parameterization is according to equation (\ref{eqn:schechter}).}
\tablewidth{0pt}
\tablehead{z	 & log$_{10}$($\phi_{*}$) [Mpc$^{-3}$]	& log$_{10}$($\phi_{*}$) [Mpc$^{-3}$] Error &	log$_{10}$$(M^*/M_{\odot})$	&		log$_{10}(M^*/M_{\odot})$ Error	&	$\alpha_s$	&	$\alpha_s$ Error}

0.0&	-4.27	&0.05	&10.76	&0.03	&-1.46	&0.02\\
0.5&	-4.53	&0.05	&10.85	&0.031	&-1.55	&0.018\\
1.0&	-4.52	&0.04	&10.76	&0.026	&-1.52	&0.018\\
1.5&	-4.79	&0.24	&10.89	&0.423	&-1.56	&0.022\\
2.0&	-5.21	&0.05	&10.93	&0.036	&-1.61	&0.018\\
2.5&	-5.25	&0.04	&10.62	&0.021	&-1.58	&0.02\\
3.0&	-5.61	&0.05	&10.67	&0.028	&-1.6	&0.019\\
3.5&	-6.03	&0.04	&10.63	&0.022	&-1.68	&0.019\\
4.0&	-6.02	&0.04	&10.33	&0.016	&-1.61	&0.021\\
4.5&	-6.09	&0.03	&10.14	&0.009	&-1.52	&0.021\\
5.0&	-7.27	&0.09	&10.45	&0.051	&-1.83	&0.03\\
5.5&	-7.54	&0.41	&10.08	&0.426	&-1.88	&0.059\\
6.0&	-9.3	&0.18	&11.02	&0.458	&-1.96	&0.076\\

\label{tab:smf_blue_satellites_C}
\end{deluxetable*}

\begin{deluxetable*}{rrrrrrr}
\tabletypesize{\scriptsize}
\tablecaption{This table lists Schechter fits of the sample of Model C for red satellites. Parameterization is according to equation (\ref{eqn:schechter}).}
\tablewidth{0pt}
\tablehead{z	 & log$_{10}$($\phi_{*}$) [Mpc$^{-3}$]	& log$_{10}$($\phi_{*}$) [Mpc$^{-3}$] Error &	log$_{10}$$(M^*/M_{\odot})$	&		log$_{10}(M^*/M_{\odot})$ Error	&	$\alpha_s$	&	$\alpha_s$ Error}

0.0&	-4.93	&0.27	&10.88	&0.04	&-1.69	&0.103\\
0.5&	-4.85	&0.28	&10.86	&0.037	&-1.67	&0.111\\
1.0&	-5.14	&0.26	&10.86	&0.038	&-1.71	&0.104\\
1.5&	-5.23	&0.23	&10.85	&0.033	&-1.69	&0.09\\
2.0&	-5.48	&0.25	&10.82	&0.032	&-1.71	&0.094\\
2.5&	-5.98	&0.18	&10.82	&0.032	&-1.81	&0.066\\
3.0&	-6.95	&0.29	&10.93	&0.041	&-2.04	&0.094\\
3.5&	-6.94	&0.25	&10.81	&0.039	&-1.94	&0.103\\
4.0&	-7.26	&0.25	&10.7	&0.046	&-1.99	&0.111\\
4.5&	-8.02	&0.39	&10.81	&0.054	&-2.14	&0.147\\
5.0&	-8.3	&1.05	&10.57	&0.084	&-2.15	&0.164\\
\label{tab:smf_red_satellites_C}
\end{deluxetable*}

\subsection{SHMR at $z=0$} \label{ap:SHMR}
In Table \ref{tab:sh} the values of the SHMR is plotted for different mass bins in the range of $10^{10}M_{\odot}$ and $10^{14}M_{\odot}$ in halo mass. These values have been calculated by binning over 0.5 dex in halo mass. The halo mass given in the table is the mean halo mass of the sample being binned over. Bins without values did not consist of at least two galaxies of the specific type within our sample.

\begin{deluxetable*}{lrrrrrr}
\tabletypesize{\scriptsize}
\tablecaption{This table lists the values of the SHMR at $z=0$ for all centrals, red centrals and blue centrals and its corresponding scatter for Model A. The figure with all individual points is given in Figure \ref{fig:sh_scatter}.}
\tablewidth{0pt}
\tablehead{log$_{10}(M_h/M_{\odot})$	 & log$_{10}(M_s/M_h)$ all & $\sigma$ (dex) all &	log$_{10}(M_s/M_h)$ blue 	& $\sigma$ (dex) blue 	& log$_{10}(M_s/M_h)$ red & $\sigma$ (dex) red}
10  & -2.98 &	0.09	&	-2.98 &	0.09 &	- &	- \\
10.5 & -2.9 & 0.14	&	-2.90 &	0.14 & 	- &	- \\
11 & -2.59 & 0.24	&	-2.59 &	0.24 &	-2.93 &	0.30\\
11.5 & -2.12 & 0.27	&	-2.11 &	0.26 &	-2.34 &	0.38\\
12 & -1.89 & 0.30	&	-1.80 &	0.20 &	-2.07	&	0.37\\
12.5 & -2.01 & 	0.36 &	-1.68 &	0.15 &	-2.10 &	0.35\\
13 & -2.30 & 0.33	&	-1.72 &	0.14 &	-2.31	&	0.33\\
13.5 & -2.56 & 0.24	&	- &	- &	-2.56 &	0.24\\
14 & -2.84 &	0.19 &	- &	 - &	-2.84 &	0.19\\
\tablecomments{(-) Symbols are indicating that our sample did not generate any objects of the specific kind in the specified mass range.}
\label{tab:sh}
\end{deluxetable*}

\end{document}